# Automation and Reuse Practices in GitHub Actions Workflows: A Practitioner's Perspective

HASSAN ONSORI DELICHEH, University of Mons, Belgium

GUILLAUME CARDOEN, University of Mons, Belgium

ALEXANDRE DECAN*, University of Mons, Belgium

TOM MENS, University of Mons, Belgium

GitHub natively supports workflow automation through GitHub Actions. Yet, workflow maintenance is often considered a burden for software developers, who frequently face difficulties in writing, testing, debugging, and maintaining workflows. Little knowledge exists concerning the automation and reuse practices favoured by workflow practitioners. We therefore surveyed 419 practitioners to elucidate good and bad workflow development practices and to identify opportunities for supporting workflow maintenance. Specifically, we investigate the tasks that practitioners tend to automate using GitHub Actions, their preferred workflow creation mechanisms, and the non-functional characteristics they prioritise. We also examine the practices and challenges associated with GitHub's workflow reuse mechanisms. We observe a tendency to focus automation efforts on core CI/CD tasks, with less emphasis on crucial areas like security analysis and performance monitoring. Practitioners strongly rely on reusable Actions, but reusable workflows see less frequent adoption. Furthermore, we observed challenges with Action versioning and maintenance. Copy-pasting remains a common practice to have more control and avoid the complexity of depending on reusable components. These insights suggest the need for improved tooling, enhanced support for a wide range of automation tasks, and better mechanisms for discovering, managing, and trusting reusable workflow components.

## 1 Introduction

GitHub is the leading platform for collaborative software development and cloud hosting of public or open-source software (OSS) projects. According to GitHub's 2025 Octoverse report [27], more than one billion contributions were made on GitHub to public repositories and repositories with a license accepted by the Open Source Initiative.[1] Driven

---

Authors' Contact Information: Hassan Onsori Delicheh, hassan.onsoridelicheh@umons.ac.be, University of Mons, Mons, Belgium; Guillaume Cardoen, University of Mons, Mons, Belgium, guillaume.cardoen@umons.ac.be; Alexandre Decan, University of Mons, Mons, Belgium, alexandre.decan@umons.ac.be; Tom Mens, University of Mons, Mons, Belgium, tom.mens@umons.ac.be.

by the need for rapid and consistent software releases of high quality, the automation of repetitive development tasks (e.g., testing, compilation, building, quality analysis, security analysis, packaging, releasing, and deploying software) has become necessary. To this end, automation workflows have become increasingly essential [25]. Specifically, Continuous Integration, Deployment, and Delivery (CI/CD) have emerged as pivotal practices within collaborative software development and the broader DevOps movement. They have been shown to improve the development process through more rapid releases, more effective pull request (PR) merging, early bug detection, and higher code quality [31, 63, 70]. While a wide range of CI/CD tools and services have served developers for many years, the move towards cloud-based CI environments with deep integration of CI/CD services has fundamentally reshaped how developers automate their workflows [28]. It allows developers to free up local machine resources by running workflows remotely, contributing to more streamlined collaborative development cycles, simplifying the overall integration process, and enabling more frequent deployments.

In November 2019, GitHub publicly released GitHub Actions,[2] a cloud-native workflow automation tool integrating direct CI/CD capabilities within its platform. This empowered GitHub repository maintainers to streamline their workflow automation processes directly within their repositories. Soon after its official release, GitHub Actions became the dominant workflow automation tool for GitHub repositories [28]. It enables developers to define workflow configuration files to automate various tasks, such as building, testing, releasing and deploying code. Workflow runs can be triggered by specific events, such as code pushes, PRs, or schedules, facilitating their seamless integration in the development process. Workflows can also rely on reusable Actions and reusable workflows, which can be shared and reused across repositories. These reuse mechanisms enhance efficiency in automating software development tasks.

Despite the widespread use of GitHub Actions, the practices and priorities of workflow developers and maintainers with respect to automation and especially reuse have not been thoroughly explored in empirical research. Developing a deeper understanding of how GitHub Actions workflow maintainers approach automation tasks, adopt the available reuse mechanisms, and balance important non-functional qualities, is essential for advancing best practices and refining tool support. Building upon prior work, this study adopts a practitioner's perspective to closely examine the non-functional priorities that drive workflow development and maintenance activities, as well as the awareness, motivations, and challenges encountered by maintainers with the available reuse mechanisms. To this end, we conducted a survey involving 419 GitHub workflow practitioners on these key aspects. More specifically, we focus on two main goals:

$G_1$: *Understanding the automation practices adopted by workflow developers.* The increasing adoption of GitHub Actions for automating software repository workflows to reduce manual effort [11] emphasises the need for a deeper understanding of the tasks being automated, the mechanisms used to develop workflows, and the non-functional qualities that drive their usage. Our findings reveal that practitioners mainly automate core software development tasks such as building and testing, while security, performance monitoring, and documentation get less attention. Reliability is prioritised over security and reusability. Workflows are mostly created from scratch or based on existing ones, while the use of starter templates or workflow generation tools remains limited.

$G_2$: *Understanding the reuse practices followed by workflow developers.* While GitHub Actions proposes several reuse mechanisms, it is yet poorly understood whether and how they are used by workflow maintainers, what are the motivations for (not) using them, what are the characteristics that influence their usage, which challenges maintainers face when incorporating them into workflows, and what are the consequences of using them (e.g., in terms of workflow quality and security). Our findings reveal that copy-pasting from workflows is a common strategy, motivated by its

---

convenience and flexibility. Workflow maintainers extensively reuse Actions shared by others but seldom adopt reusable workflows. Maintainers also struggle with the discoverability of suitable reusable artefacts, their maintainability, and issues related to outdated dependencies.

Combined together, the findings of both goals provide qualitative insights into the good and bad practices of workflow maintenance and reuse. They underscore the need for increased awareness and more reliable support for maintaining workflow reliability, quality, and security. They also call for better automated support for workflow creation, debugging, and optimisation. With respect to reuse, there is a need for better mechanisms for versioning, sharing, and discovering reusable artefacts. This should be combined with enhanced documentation and awareness to promote more reliable workflow reuse practices.

The remainder of this article is structured as follows. Section 2 presents the related work on qualitative CI/CD studies and empirical research on GitHub Actions. Section 3 reports on the methodology that was followed to conduct the survey and analyse its results. Section 4 presents an analysis of the results pertaining to goal $G_1$, focusing on workflow automation practices. Section 5 presents an analysis of the results pertaining to goal $G_2$, focusing on workflow reuse practices. Section 6 reports on the threats to validity of our research, and Section 7 concludes.

## 2 Related work

### 2.1 Qualitative studies on CI/CD practices and principles

Some early qualitative studies explored CI practices within industrial contexts. Ståhl and Bosch [59] undertook a multi-case study at Ericsson AB to validate the purported benefits of CI claimed in literature. Through in-depth interviews with various stakeholders, they revealed that CI significantly facilitated parallel development, leading to tangible increases in developer productivity. Furthermore, they demonstrated that CI enhanced project predictability by enabling the early detection of issues and facilitating early non-functional testing. The study also highlighted a positive impact on communication, extending beyond individual teams to improve collaboration across larger project structures.

Expanding beyond industrial settings, subsequent research broadened its scope to encompass OSS projects and the diverse practices of developers within these communities. Hilton et al. [31] conducted a mixed-methods empirical analysis of CI usage in OSS projects to gain a holistic understanding. The study focused on three areas: they analysed 34,544 GitHub projects to establish which CI systems were being used, investigated 1,529,291 builds to understand usage patterns, and surveyed 442 developers to find out why they were (or were not) adopting CI. This multi-faceted investigation provided valuable insights into the usage patterns, associated costs, and perceived benefits of CI in OSS projects. Embury and Page [24] examined the pedagogical aspects of CI, focusing on the impact on the build health of projects by undergraduate students in software engineering courses. They concluded that providing students with access to CI results significantly improved their ability to manage build quality, particularly in areas directly related to assessment criteria.

As CI practices matured, identifying and mitigating CI anti-patterns became a central focus of research. Vassallo et al. [64] developed the CI-Odor tool to detect anti-patterns through the analysis of build logs and repository data. This tool was validated through surveys with developers, confirming its effectiveness in identifying issues that diminish the benefits of CI. Zampetti et al. [68] compiled a comprehensive catalogue of CI bad smells through expert interviews and analysis of Stack Overflow posts, and validated this catalog through a survey of professional developers, providing a deeper understanding of the challenges encountered during CI implementation. Kochhar et al. [38] explored the impact on CI practices of organisational transitions in Microsoft from closed to open source systems. Their study highlighted the influence of open-sourcing on CI and other development processes, providing insights from both the company's and the OSS community's perspectives.

Helis et al. [30] analysed 162K+ PRs in 87 GitHub projects and 450 responses of a survey to investigate the relationship between CI and software delivery speed in OSS projects. They revealed that, while CI does not necessarily accelerate the merging of PRs, it significantly improves decision-making on submissions. This emphasises the crucial role of CI in enhancing contributor engagement and confidence. Santos et al. [55] explored the monitoring of CI practices, revealing a gap between developer needs and the support provided by CI services. Through document analysis and developer surveys across 121 OSS projects, they identified that test coverage was frequently discussed, but other crucial practices like build health and time to fix a broken build were often overlooked. Their findings indicated that developers desired more comprehensive CI monitoring, yet existing CI services offered limited native support, necessitating reliance on cumbersome third-party tools. Consequently, the study concluded that CI monitoring is generally underutilised, highlighting the need for CI services to integrate robust monitoring capabilities to encourage and facilitate better development practices.

## 2.2 Empirical research on GitHub Actions

Focusing on GitHub Actions specifically, empirical research has explored its adoption, its impact on development practices, and associated challenges related to security and dependency management. Initial studies focused on GitHub Actions' growing prominence in the CI/CD landscape. Golzadeh et al. [28] performed a quantitative analysis of 91,810 GitHub repositories to study how developers change their CI/CD environment over time. They noticed that the introduction of GitHub Actions coincided with a substantial reduction in the use of other CI/CD services, such as Travis, CircleCI, and Azure. Rostami Mazrae et al. [54] investigated the rationale behind the usage, co-usage, and migration between 31 distinct CI/CD tools. Their study, involving in-depth interviews with 22 seasoned software professionals, revealed a clear trend of migration towards GitHub Actions and pinpointed the primary drivers behind this migration. The core identified motivations for CI usage were enhancing reliability, productivity, security, and speed, while simultaneously reducing cost and effort.

Decan et al. [21] provided an overview of the use of GitHub Actions workflows based on a dataset of 67K+ GitHub repositories. They observed that nearly all repositories using workflows rely on reusable Actions, and therefore advocated for more in-depth studies on the benefits, challenges and impact of this reuse mechanism. Saroar et al. [56] mined the GitHub Marketplace as a software production platform, providing an analysis of 440 Apps and 7,878 Actions distributed through the Marketplace. Comparing these tools with academic production tools reported in the research literature, they observed that practitioners often use automation tools for tasks belonging to the categories of continuous integration and utilities, while researchers tend to focus more on code quality and testing.

Kinsman et al. [37] investigated changes in various development activity indicators related to adopting GitHub Actions by examining 3,190 GitHub repositories. They revealed that this adoption led to an increased monthly number of rejected PRs, alongside a decrease in the number of commits on successfully merged PRs. Wessel et al. [66] corroborated and extended these findings by observing that the adoption of GitHub Actions leads to more PR rejections and an increased PR acceptance time. Building upon this, Chen et al. [12] explored the relationship between GitHub Actions usage and repository features. Utilising statistical models, they analysed how GitHub Actions usage impacts the frequency of changes and the efficiency of resolving PRs and issues. These insights aid repository maintainers in understanding and mitigating potential negative consequences of GitHub Actions adoption.

More recent studies on the lifecycle of GitHub Actions workflows revealed important trends and challenges. Through a manual study of workflow evolution in 10 popular GitHub repositories, Valenzuela-Toledo and Bergel [60] revealed that bug fixes and CI/CD improvements drive workflow maintenance. They highlighted the hidden costs of automation, advocating for better resource planning, documenting best practices, and improving tools for dependency tracking and error reporting. Valenzuela-Toledo et al. [61] empirically investigated workflow maintenance across 183 GitHub repositories. Their study demonstrates that, while automation via GitHub Actions streamlines CI/CD practices, it also introduces ongoing and sometimes overlooked maintenance needs. Workflow file changes are typically loosely connected to source code changes, with many updates arising from improvements to CI/CD pipelines, infrastructure management, bug resolution, and documentation. These tasks are most often carried out by a small group of core contributors, who are also frequently involved in production and test code maintenance. The research highlights hidden maintenance costs accompanying workflow automation and calls for best-practice guidelines and supporting tools for sustained workflow maintenance, with particular emphasis on dependency tracking and error reporting. To further explore this hidden cost and how to overcome it, Bouzenia and Pradel [10] conducted an empirical study of resource usage in 1.3 million workflow runs from 952 GitHub repositories. They observed that over 91% of the resources are consumed by testing and building, and that existing effective optimisations such as caching remain significantly underutilised. They demonstrated that simple platform changes could yield substantial resource savings, such as cutting runner time by up to 31%. Complementing this research, Valenzuela-Toledo et al. [62] focused on the challenge of diagnosing workflow failures. They explored the potential of Large Language Models (LLMs) to generate contextual descriptions for workflow failures, finding that while LLMs can assist developers in understanding common errors, improved reasoning capabilities are still required for complex CI/CD scenarios. Zheng et al. [71] conducted an empirical study on workflow failures, manually analysing 375 failed executions and surveying 151 developers. They proposed a taxonomy of 16 distinct failure types, finding that test issues, compilation problems, and configuration errors are the most common failure causes. For resolving these failures, they identified debugging and resource limits as the key challenges.

Ghaleb et al. [26] empirically studied the use of various CI/CD services (including GitHub Actions) in 2,557 Android App repositories. They found that configuration files are often complex and lack standardisation, with most projects focusing primarily on the initial setup and build processes. Testing activities were present in roughly half of the projects, predominantly implemented through standard unit tests, while deployment was comparatively less common and observed in only a minority of cases. Furthermore, the maintenance of these workflows typically occurred on a bi-monthly basis, particularly in repositories utilising GitHub Actions, where the effort mainly involved routine fixes and updates. Zhang et al. [69] analysed 6,590 Stack Overflow questions and 315 GitHub issues, resulting in a taxonomy of GitHub Actions workflow challenges. These challenges range from initial setup and configuration to maintenance and operation of workflows, encompassing areas such as configuration, dependency management, triggering, monitoring, security management, resource sharing, containerised services, and deployment. They offered 56 specific and actionable solution strategies for the identified challenges, serving as a practical guide for troubleshooting and optimising GitHub Actions workflows. Khatami et al. [36] identified workflow smells that are indicative of underlying issues in GitHub Actions. Through mining and manual analysis of frequent change patterns in 83 repositories, they proposed 22 potential smells, of which seven were subsequently confirmed via PRs and maintainer feedback across 32 repositories.

Mastropaolo et al. [46] focused on auto-completion of GitHub Actions workflows. They introduced GH-WCOM, a transformer-based method to autonomously suggest workflow completions, for example by populating a workflow *job* based on its textual documentation, or by suggesting the next *step* to be added to a *job*. Huang and Lin [32] introduced

CIGAR, a contrastive transformer-based learning model, to discover and select appropriate Actions for workflows. Trained on thousands of Actions, the model significantly improves upon recommending relevant Actions for workflow tasks compared to a simple keyword-based search on the GitHub Marketplace. More specifically, CIGAR is specifically designed to understand the intent behind the name descriptions of workflow steps (for example, "setup Python", "upload artefact", or "deploy application") and match them to the most suitable Actions. Nguyen et al. [49] introduced a technique leveraging transformer and few-shot learning models to address the issue of unclear or missing categorisations of Actions on the GitHub Marketplace. Their technique automatically categorises Actions based on their README files.

Security is a significant concern for GitHub Actions workflows. Their substantial reliance on reusable Actions [20] introduces potential vulnerabilities if these Actions are compromised or contain malicious code. Benedetti et al. [7] proposed a security evaluation framework and discovered numerous security issues in GitHub Actions workflows within OSS projects. Koishybayev et al. [39] identified prevalent risks such as excessive permissions, code execution vulnerabilities, and the use of unverified Actions, further underscoring the software supply chain risks inherent in GitHub Actions workflows. Onsori Delicheh and Mens [52] recommended mitigation strategies specifically targeting security issues in GitHub Actions workflows that are related to software supply chain attacks and workflow misconfigurations.

Dependency-related issues have also been a focus of research. Decan et al. [20] quantified the prevalence of outdated reusable Actions, highlighting the challenge of maintaining up-to-date dependencies. Onsori Delicheh et al. [51] studied the dependency structure of reusable Actions, revealing a strong reliance on JavaScript packages and complex transitive dependencies, raising maintainability concerns. They also explored security vulnerabilities stemming from these dependencies, finding widespread issues and highlighting the need for improved security measures [22].

Focusing on the development and use of reusable Actions, Saroar and Nayebi [57] surveyed 25 workflow maintainers and 44 Action developers to investigate their motivations, preferred practices, and principal challenges when creating, integrating, and troubleshooting Actions. The surveyed workflow maintainers favoured Actions produced by verified authors and those with a considerable number of stars, particularly in presence of multiple suitable Actions of comparable quality. When confronted with software defects or limited documentation, maintainers often migrate to alternative Actions. Debugging GitHub Actions workflow files was identified as an important challenge, reflecting broader limitations in diagnostic support for GitHub Actions. While the work by Saroar and Nayebi is closely aligned with ours, its focus and research questions are distinct. Saroar and Nayebi focus specifically on the reusable Actions, primarily (though not exclusively) surveying Action developers (44 respondents) to identify their motivations, selection criteria, and barriers associated with creating, publishing, and adopting Actions. In contrast, our work targets a large number of workflow maintainers (419 respondents), aiming to understand the types of tasks automated by GitHub Actions workflows, the approaches employed in constructing these workflows, and the non-functional qualities prioritised throughout development and maintenance. Furthermore, our survey provides a considerably broader and more detailed perspective on the challenges of workflow reuse, encompassing traditional copy-and-paste strategies as well as the adoption of various reuse mechanisms such as Actions, composite Actions and reusable workflows.

Chomątek et al. [13] used LLMs to perform a large-scale analysis of CI/CD services (including GitHub Actions) across 28,770 GitHub repositories. Their findings confirm that core practices are near-universal, with building and testing tasks being almost universally adopted. In contrast, specialised practices saw much lower adoption, with deployment and artefact creation being moderately implemented, and practices like Static Application Security Testing (SAST) and containerisation remaining significantly underutilised. Their results also highlight different adoption patterns among distinct CI/CD services, revealing that GitHub Actions leads in the adoption of practices such as linting and SAST compared to other platforms.

We will discuss the relation between our results and these prior works in detail in Section 4.4 and Section 5.5.

## 3 Methodology

This section presents the methodology we followed to construct and conduct the online survey. Section 3.1 describes the design and validation of the questionnaire. Section 3.2 explains how we select the respondents. Section 3.3 reports on how we contacted them, how we collected survey data, what are the data protection measures taken and how we ensured to retain only responses that were relevant for further analysis. Finally, Section 3.4 explains the mixed-methods approach that was followed to process the responses.

### 3.1 Questionnaire design

The questionnaire was collaboratively designed with the aim of exploring the two primary research goals: ($G_1$) understanding the automation practices adopted by workflow developers; and ($G_2$) understanding the reuse practices followed when creating and maintaining workflows. The questionnaire starts by outlining the research objectives, followed by an initial question to assess respondents' familiarity with GitHub Actions and with other CI/CD tools in general. The remainder of the questionnaire is divided into two sections, each targeting a specific research goal. The first section focuses on $G_1$ and comprises three questions related to the tasks being automated by workflows, the mechanisms employed to create these workflows, and the importance of specific non-functional characteristics during workflow maintenance. The second section focuses on $G_2$ and comprises four questions to understand the motivations behind reuse in workflows, the characteristics that influence the selection and usage of reusable components, and the challenges faced when incorporating them into workflows.

The questionnaire was designed to be concise, with the aim of being completed within 15 minutes. This was achieved by limiting the number of questions and ensuring that each of them was directly related to the research goals. Following established design guidelines [41], the questionnaire was created from the respondents' perspective, using a familiar language, with precise instructions and examples provided where necessary. This was done to ensure that respondents could easily understand the questions and provide accurate answers without requiring additional clarification.

The options presented in each question were agreed upon based on a preliminary analysis of existing literature, discussions with domain experts and several rounds of discussions between the co-authors. Free-text fields were included to allow respondents to provide insights on additional aspects that might have been overlooked when designing the survey. Many questions targeted the perceived *importance* or *frequency* of various aspects related to workflow automation and reuse. These questions were formulated using a four-point Likert scale, omitting a neutral choice to avoid passive neutrality and elicit decisive responses that are more suitable data for quantitative analysis [9]. An "*I don't know*" choice was included wherever appropriate to prevent respondents from making arbitrary choices when they were uncertain or lacked the necessary knowledge. For those questions where more than one option was relevant, multiple-choice questions with checkboxes were provided to allow respondents to select multiple applicable options from predefined lists.

The questionnaire underwent multiple iterations among all co-authors. An initial version was validated through several brainstorm meetings to ensure that all pertinent elements for reaching the research goals were included. This version underwent multiple iterations based on mutual agreement to review the wording, sequencing, and coverage. We also solicited and incorporated feedback from domain experts in CI/CD and GitHub Actions, which led to minor adjustments to reduce ambiguity and avoid redundancy. Once all co-authors agreed that no further changes were necessary, a pilot study was conducted with four software practitioners experienced in maintaining GitHub Actions workflows. Each practitioner completed the survey in the presence of (at least) two authors, allowing immediate feedback on the length, duration, understandability, and clarity of the questionnaire. Any remaining unclarity or ambiguity was removed, resulting in the final version of the questionnaire, that can be found on Zenodo.[3]

### 3.2 Target audience

To achieve the research goals, a primary requirement was to target respondents involved in GitHub Actions workflow maintenance, ensuring the collection of responses from practitioners with demonstrable familiarity in workflow creation and maintenance. This was considered crucial for receiving informed and insightful responses. We aimed to reach a sample size of at least 384 fully completed responses, in order to ensure a 95% confidence level with a 5% margin of error. This is a widely accepted standard in qualitative empirical studies for reaching statistical significance [16], and allows to generalise the findings to the broader population of GitHub Actions workflow practitioners with a reasonable degree of accuracy [14].

In order to identify practitioners involved in maintaining GitHub Actions workflows in GitHub repositories, we need a dataset containing information on contributors that are known to have made changes to workflow files. A comprehensive dataset providing such information was created by Cardoen et al. [11]. It provides the history of workflow files, including metadata on their commit history, such as the committer's login name and email address. We relied on version 2024-10-25[4] of this dataset, containing 219K+ workflow file histories from more than 43K public software development repositories on GitHub, and excluding repositories that did not reflect active software development. We used the workflow file commit histories to identify individuals who had actively contributed to the development and maintenance of GitHub Actions workflows. To ensure we target individuals with recent involvement in workflow maintenance, we only selected those who had committed ten changes to workflow files with at least one of these commits made during the last three months prior to the data collection date. By applying these filters, we identified 6,500 potential respondents (i.e., around 17 times the number of respondents required to reach the desired sample size).

### 3.3 Data collection

The survey was conducted by researchers of UMONS within the framework of the university's public interest research mission, with research benefits to the scientific community as well as to the open-source software development community at large. Prior to conducting the survey, the researchers created a Data Management Plan (ID 210808) that was validated by the university's Data Production Officer. The survey was conducted using LimeSurvey,[5] a popular open-source survey tool installed on a secured university-hosted server. We employed an online survey, a method chosen for its accessibility and ability to reach a broad audience of contributors [41]. In accordance with GDPR,[6] the survey was based on auto-generated tokens to decouple responses from the respondent names or email addresses. Anonymised responses were analysed confidentially by authorised researchers from the Software Engineering Lab. No personal information or demographics were asked during the survey, except for the option offered to respondents to leave an email address of their choice at the end of the survey, in order to be informed of the outcome of the survey.

To limit unsolicited communication, respondents were contacted only once, and participation was entirely voluntary. We did not send any reminder emails. To improve the survey participation rate, we followed some recommendations provided by Smith et al. [58]. More specifically, we designed the survey to take not more than 15 minutes to complete,

---

and informed respondents about this. We sent individual personalised emails to each respondent, and we increased credibility by explicitly mentioning the official affiliations and titles of the involved researchers. We also provided the necessary information pertaining to the data protection policy.

Initially, we contacted respondents in batches of 100 randomly selected individuals from a pool of 6,500 individuals to gather quick feedback regarding potential issues with the online survey tool, as well as to identify any unforeseen challenges. This approach allowed us to refine our communication strategy and immediately address some minor concerns raised by the respondents. Since no technical concerns were reported, we increased the batch size to 500 invitations per day. We reached the desired sample size of 384 responses after having sent out 3,500 invitations, and stopped sending out more invitations at that point. The entire process of sending invitations and receiving responses took place from 23 January 2025 to 6 March 2025.

Out of the 3,500 invited practitioners, 592 opened the survey questionnaire and provided responses to the questions. However, not all respondents fully completed the questionnaire. We ultimately analysed 419 complete responses, representing an 11.97% response rate. On average, participants spent 10 minutes and 23 seconds completing the survey, with a median time of 8 minutes and 15 seconds.

Table 1. Familiarity of respondents with GitHub Actions and other CI/CD tools.

| GitHub Actions → ↓ other CI/CD tools | not familiar at all | slightly | moderately | very familiar | total |
|---|---|---|---|---|---|
| not familiar at all | 0.0% | 0.5% | 3.8% | 2.9% | **7.2%** |
| slightly | 0.0% | 1.2% | 13.4% | 14.1% | **28.6%** |
| moderately | 0.0% | 0.2% | 12.4% | 26.7% | **39.4%** |
| very familiar | 0.0% | 0.2% | 2.9% | 21.7% | **24.8%** |
| **total** | **0.0%** | **2.2%** | **32.5%** | **65.4%** | **100%** |

The first question of the survey aimed to ascertain the respondents' level of familiarity with CI/CD tools in general and with GitHub Actions specifically. Respondents who indicated no familiarity at all with GitHub Actions were prevented from continuing the questionnaire. Table 1 reports on this familiarity, confirming that all 419 respondents demonstrated at least some familiarity with GitHub Actions workflows, and nearly all of them (97.9% = 32.5% + 65.4%) reported to be *moderately* or *very familiar* with GitHub Actions. Only 2.2% of the respondents, corresponding to 9 participants, reported being only *slightly familiar* with GitHub Actions. This confirms that the respondents are sufficiently familiar with GitHub Actions to answer the survey questions that focus on automation and reuse practices in GitHub Actions workflows. For comparison, the familiarity with other CI/CD tools is slightly lower, with 7.2% of the respondents indicating that they were *not familiar at all* with other CI/CD tools, and 28.6% being only *slightly familiar* with them. The final dataset of anonymised responses is provided on Zenodo.[7]

### 3.4 Data analysis

Section 4 and Section 5 will present the survey results and the analysis of the responses undertaken to address the research goals. Before doing so, we describe the data processing and analysis methods we used to extract insights from the collected responses. The code used for these quantitative and qualitative analyses is available in a replication package on Zenodo.[8]

---

[7]See https://doi.org/10.5281/zenodo.15422634.
[8]See https://doi.org/10.5281/zenodo.15422987.

We started by performing descriptive statistics on the survey responses, mostly consisting of ordinal labels (derived from the Likert scales) related to frequency and importance, but also including responses to multiple-choice questions (answered through checkboxes). For the ordinal data, we calculated frequencies and percentages to understand the distribution of responses across the various categories of options. For the multiple-choice questions, we calculated the frequency of selecting each option, providing insights into the prevalence of each chosen response. To gain a deeper understanding of the relationships between items in the survey responses, we employed the well-known Apriori algorithm to identify frequent itemsets [2]. While descriptive statistics provide an overview of individual items, frequent itemset analysis enables to uncover combinations of items that are frequently considered together.

After the descriptive analysis, we performed inferential statistics to explore relationships and differences within the survey data. For each question based on a Likert scale, we determined the relative order of option selection (e.g., which tasks are more often performed with GitHub Actions than others). To achieve this, ordinal labels were first transformed into numerical values on a linear scale. Then, in order to identify statistically significant differences, we relied on non-parametric Kruskal-Wallis tests [17]. This test is suitable for comparing multiple groups (i.e., selected options) without assuming normality or other parametric test conditions. For the questions where the null hypothesis could be rejected (i.e., indicating that at least one option was selected significantly more than others), we proceeded with post-hoc Dunn's tests [23] to identify specific pairs of options whose values exhibit a statistically significant difference (e.g., to identify that the tasks of "testing" and "deployment" have a statistically different frequency). While Dunn's test indicates whether groups differ, it does not specify which group is higher. We therefore computed the corresponding Mann-Whitney U statistics [45] (i.e., sum of ranks) to determine the direction of the difference (e.g., to determine that "testing" is more frequently performed than "deployment"). This resulted in a partial order on the groups, indicating which specific options or responses were favoured over others by the population of respondents.

The $p$-values of Kruskal-Wallis' and Dunn's tests were used to determine statistical significance at a level of $\alpha = 0.05$. Given that we performed 238 statistical tests (essentially corresponding to one Kruskal-Wallis test per question and subsequent Dunn's tests for each pair of options within those questions), we applied a Benjamini-Hochberg correction on the $p$-values to control the family-wise error rate and minimise the risk of false positives [8]. This correction method is widely accepted for controlling the false discovery rate in multiple testing scenarios, in order to maintain a balance between statistical power and the risk of false positives [29]. After correction, 208 out of 238 tests were statistically significant (i.e., their adjusted $p$-values were below the $\alpha$ level).

To process the detailed feedback and perspectives provided by respondents in the free-text fields, we conducted a qualitative analysis following the guidelines of Corbin and Strauss [18]. An open coding process was initially employed, to segment the data and label it with meaningful descriptors. Subsequently, axial coding was utilised to group these codes into broader categories based on thematic similarities and differences. To ensure rigour and mitigate researcher bias, two authors independently conducted this coding process. Following this, the authors discussed together to compare their analyses. In case of disagreements, a third author reviewed the cases and a consensus-building discussion was established to come to a mutually agreed decision. This collaborative process aimed to enhance the reliability and validity of the qualitative findings. In practice, the free text analysis did not reveal significant new insights. Participants primarily used the free-text fields to provide detailed comments and clarifications to their selection, suggesting that the pre-defined options were sufficiently comprehensive.

## 4 $G_1$: Understanding workflow automation practices

The increasing reliance on GitHub Actions for automating software development tasks highlights the need for a thorough understanding of its practical usage. Therefore, the first goal of this paper is to understand the automation practices employed by workflow developers, focusing on three key areas: (1) the tasks being automated, (2) the mechanisms used to develop workflows, and (3) the non-functional qualities that drive their usage. By analysing these elements, we seek to provide empirical insights that can inform good and bad practices and enhance the effectiveness of GitHub Actions workflows in collaborative software development.

### 4.1 Tasks being automated by workflows

The automation of repetitive development tasks through GitHub Actions workflows offers significant advantages like increased efficiency, reduced errors, and faster release cycles. To understand the variety of tasks being automated we asked respondents how frequently they use GitHub Actions workflows to automate the tasks listed in Table 2. These tasks are heavily inspired by CI activities identified in previous empirical studies (e.g., [21, 54], see Section 2).

Table 2. List of tasks being automated using GitHub Actions workflows.

| task | description |
| ---: | --- |
| testing | Automate the testing of artefacts. |
| compiling & building | Automate the compilation and building of artefacts. |
| code quality analysis | Analyse the code (statically or dynamically) for quality, compliance of coding standards, linting, code smells, … |
| version & release management | Create and publish new versions and releases, generate release notes and changelogs, ... |
| deployment | Automatically deploy applications to various environments (e.g., AWS, Azure, Google Cloud, ...). |
| documentation | Build and deploy documentation (e.g., developer or end-user documentation, websites). |
| dependency management | Automate dependency updates, scan vulnerabilities in dependencies, ... |
| security analysis | Secret management, code signing, scanning for code vulnerabilities, detecting weaknesses in repository or workflow configurations, and so on. |
| PR management | Automate PR related tasks like merging, reviewing, triaging, assigning, and closing them. |
| internal reporting | Generate internal reports or badges related to coverage results, test results, ... |
| infrastructure & config. management | Manage configurations and (cloud) infrastructures (e.g., Terraform, Cloud-Formation). |
| issue management | Automate issue-related tasks such as triaging, labeling, assigning, closing issues, and discussing issues. |
| compliance checking | Check for compliance to licenses and policies. |
| health & performance monitoring | Analyse system or network performance and related health metrics. |
| external communication | Automate with external communication channels such as email, Slack, ... |

Respondents reported using GitHub Actions workflows to automate a median of eight tasks. A median of five tasks (i.e., one third of all 15 listed tasks) are performed at least *frequently*. Figure 1 shows the reported frequency of tasks. Software **testing** stands out, with 91.2% of respondents automating it at least *frequently*. **Compiling and building** software artefacts is also highly prevalent, being performed at least *frequently* by 88.1% of respondents. **Code quality**

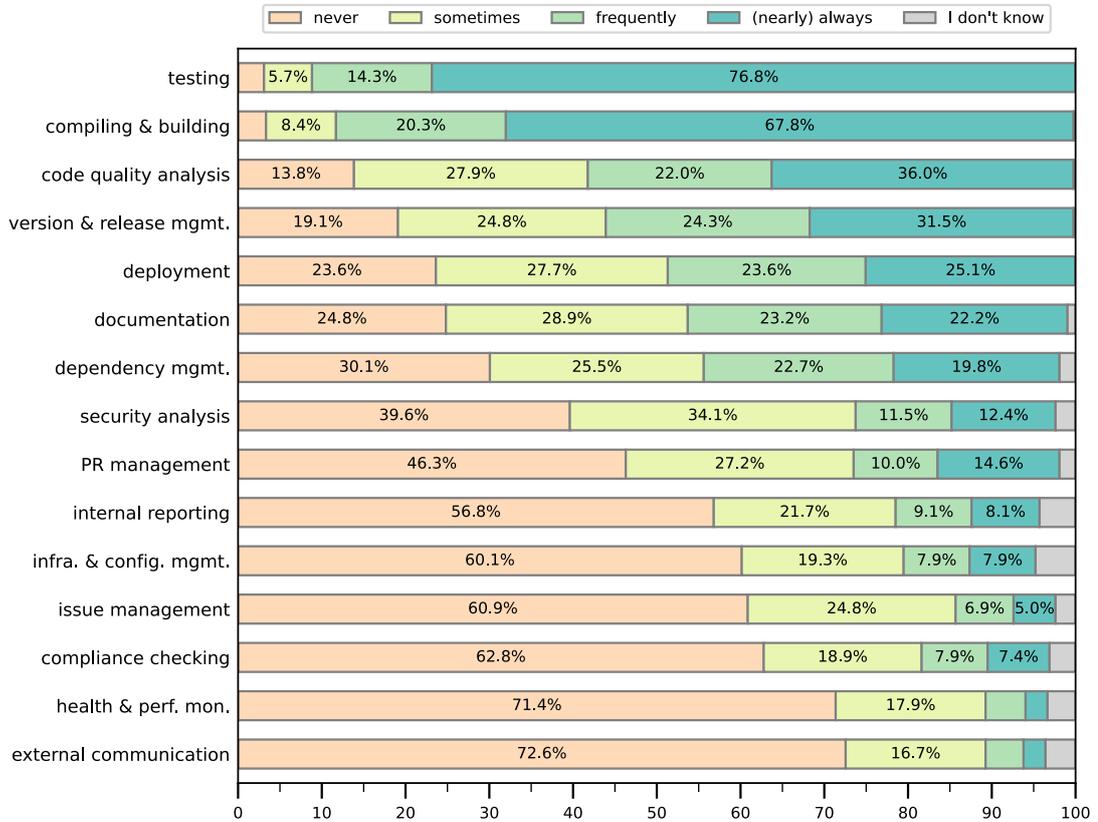

Fig. 1. Reported frequency of tasks that are being automated by GitHub Actions workflows.

**analysis** and **version and release management** are also reported to be automated at least *frequently* by the majority of respondents (58.0% and 55.8%, respectively). Three more tasks were reported to be automated at least *frequently* by at least 2 out of 5 respondents: application **deployment** (48.7%), building and deployment of **documentation** (45.4%), and **dependency management** (42.5%).

All other tasks exhibit substantially lower levels of automation, with less than one out of four respondents automating them at least *frequently*. Still, more than half of the respondents automated at least *sometimes* the tasks of **security analysis** (58.0%) and **PR management** (51.8%). Respondents reported considerably less frequently to automate at least *sometimes* the tasks of **internal reporting** (38.9%), **issue management** (36.7%), **infrastructure and configuration management** (35.1%), **compliance checking** (34.1%), **health and performance monitoring** (25.3%), and **external communication** (23.9%).

We used the Apriori algorithm (see Section 3.4) to identify frequent itemsets, i.e., combinations of items (in our case, automation tasks) that occur together frequently in the data. The analysis reveals a strong association between **testing** and **compiling and building**, appearing together in 94% of the task selections. A more comprehensive set of six tasks – including **testing**, **compiling and building**, **code quality analysis**, **version and release management**, **deployment** and **documentation** – was also frequently selected together, appearing in 49% of the task selections. These qualitative findings corroborate the quantitative observations made by Decan et al. [21]. Their analysis of all job names within 70K distinct GitHub Actions workflows revealed that the most frequently used names were `build`, `test`, `analyse`, `lint`, `release`, and `deploy`.

To further explore the relationships between frequently automated tasks, we analysed which specific tasks were favoured over others in terms of automation frequency, following the statistical analysis process outlined in Section 3.4. We visualise the results of this analysis as a partial order in Figure 2, limited to the nine most frequently automated tasks for clarity. An arrow from a task to another in this partial order indicates statistical evidence that the source task is carried out more frequently than the target task. This partial order suggests that more advanced development tasks are automated less frequently than the more essential development tasks. For example, automated compilation, building and testing is more frequent than the automation of (static or dynamic) code quality analysis.

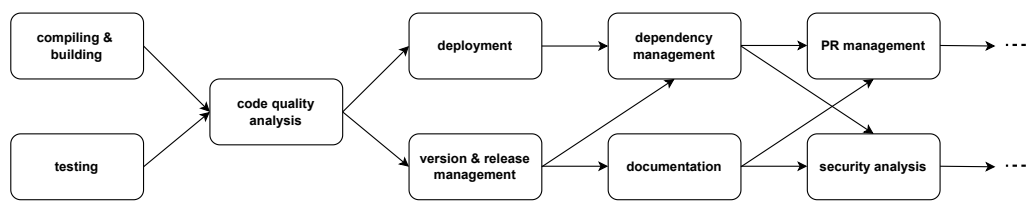

Fig. 2. Partial order of frequent workflow automation tasks, reflecting the relative choices made by respondents. The partial order is limited to the 9 most frequent tasks. The last six tasks of Figure 1 for which the majority of respondents indicated never to automate them are not shown for ease of readability.

Summary: GitHub Actions workflows are used to automate a diverse range of tasks, with development-centric CI/CD activities being the most prevalent: building and testing take the top spots, followed by code quality analysis, and then deployment alongside version and release management. Workflow maintainers are significantly less commonly automating other tasks such as internal reporting, infrastructure and configuration management, issue management, compliance checking, health and performance monitoring, and external communication.

### 4.2 Workflow creation mechanisms

Creating well-designed GitHub Actions workflows is important for achieving effective automation of tasks within repositories. Understanding how developers create these workflows can help identify common practices, potential challenges, and opportunities for improving workflow development.

Table 3 lists the different mechanisms proposed for creating workflows, using the approach explained in Section 3.1. The proposed mechanisms are based on the practical experience of the authors and domain experts, as well as the GitHub documentation.[9] Among these mechanisms, GitHub workflow templates[10] allow users to create new workflows based on predefined examples. Sometimes referred to as *starter workflows*,[11] these templates are accessible through the Actions tab in the web interface of any GitHub repository.

---

[9]https://docs.github.com/en/actions/writing-workflows
[10]https://docs.github.com/en/actions/writing-workflows/using-workflow-templates
[11]https://github.com/actions/starter-workflows

Table 3. List of mechanisms for creating new GitHub Actions workflows.

| mechanism | description |
|---|---|
| own workflow | I base myself on one of my own workflows. |
| other's workflow | I base myself on someone else's workflow. |
| from scratch | I write my workflows from scratch. |
| another source | I make use of another source (forum, Q&A platform, ...). |
| starter template | I rely on a starter workflow template suggested through the GitHub user interface. |
| other template | I rely on a workflow template provided elsewhere. |
| tool-assisted | I am assisted by a tool (e.g., CoPilot). |

We asked respondents how frequently they used the proposed mechanisms to create new workflows. Respondents reported using a median of five different workflow creation mechanisms (i.e., mechanisms used at least *sometimes*). Among those mechanisms, a median of two were used at least *frequently*. Figure 3 shows the reported frequency of using each creation mechanism. Most respondents (77.1%) created new workflows at least *frequently* by basing them on one of their **own workflows**. The large majority of respondents also reported at least *sometimes* to create new workflows **from scratch** (89.3%) or based on **other's workflow** (89.0%). A frequent itemset analysis revealed that respondents rely on multiple mechanisms to create new workflows, since 78.5% of respondents used all of the above three mechanisms at least *sometimes* for creating new workflows.

Two other creation mechanisms were less frequently used. 65.6% of respondents used **other template** at least *sometimes*, while 65.2% of respondents used **another source** at least *sometimes*. The remaining two mechanisms were even less frequent: 39.4% of respondents used **tool-assisted** mechanisms at least *sometimes*, while 37.0% did so for **starter templates**. This low frequency of using starter templates is somewhat surprising, since they are directly supported through the GitHub UI. This could be due to their limited ability to meet the needs of workflow developers. One respondent explained how their organisation built their own specific set of workflow templates: *"We have a workflow template repository in our organization on GitHub which contains common workflows used across all of our repositories."* Other respondents tried to overcome the limitations of starter templates by using custom-built automated tools to generate workflows: *"I use <HIDDEN> to create action workflows from predefined snippets"*, *"I often use <HIDDEN> to generate a GitHub Actions workflow."*, *"I use a tool which generates GHA workflows from a programming language/build system specific build definition "* and *"I use <HIDDEN> which I'm the author of."*[12] These comments suggest a clear opportunity to increase availability and configurability of GitHub's starter templates.

Based on the relative frequency reported by respondents, we computed a partial order to analyse which creation mechanisms were used more frequently than others. This partial order, shown in Figure 4, reveals a clear preference towards relying on one's **own workflow** or creating a new one **from scratch**. We hypothesised that the creation of workflows **from scratch** is associated with respondents exhibiting a high degree of familiarity with GitHub Actions. To confirm this hypothesis, we conducted a *Fisher exact test* [3] to determine whether a non-random association exists between two categorical variables. The null hypothesis $H_0$, stating that the development of workflows **from scratch** is independent of the respondent's familiarity level, was rejected yielding an odds ratio of 4.35. This signifies that respondents being *very familiar* with GitHub Actions are over four times more likely to create workflows **from scratch** compared to those being *slightly* or *moderately familiar*. These findings may be a consequence of the predominance of experienced practitioners in our survey, as explained in Section 3.3, with 65.4% of respondents being *very familiar* with GitHub Actions.

---

[12]Pointers to the tools they mentioned have been hidden as they would reveal the identity of the respondents.

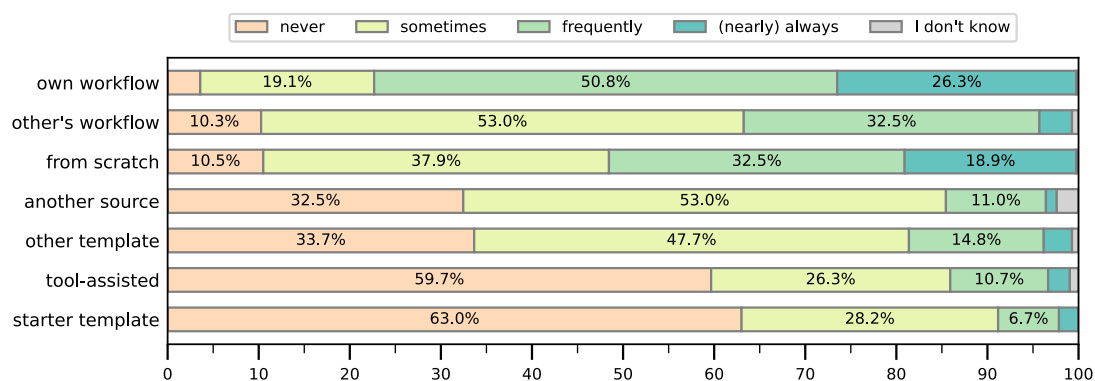

Fig. 3. Reported frequency of mechanisms used to create GitHub Actions workflows.

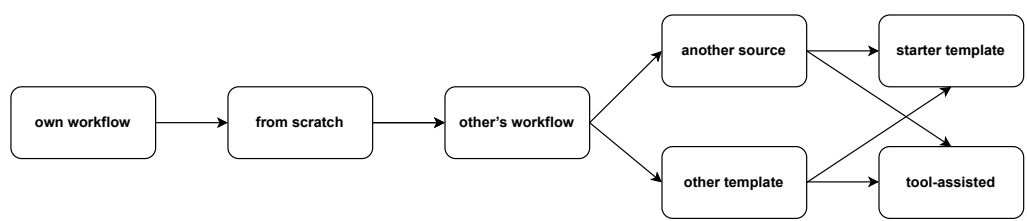

Fig. 4. Partial order of frequent mechanisms to create workflows.

SUMMARY: A large majority of respondents create new workflows either by adapting their own existing ones, by writing them entirely from scratch, or by basing them on some other workflow. Respondents who are very familiar with GitHub Actions are four times more likely to create workflows from scratch. Two mechanisms for facilitating workflow creation (tool-assisted and starter templates) see significantly less frequent use, suggesting opportunities for future improvement.

### 4.3 Importance of non-functional characteristics during workflow maintenance

In addition to ensuring that workflows fulfil the functional requirements of the tasks they automate, workflow maintainers may also value important extra-functional qualities to ensure the long-term effectiveness and sustainability of their workflows. These non-functional requirements can impact the efficiency and cost-effectiveness of ongoing adjustments and improvements [1, 65]. Table 4 lists seven considered non-functional characteristics, inspired by ISO/IEC 25010[13] and previous studies [15, 43]. Respondents were asked to indicate how important they consider each of these characteristics when maintaining workflows.

---

[13]https://iso25000.com/index.php/en/iso-25000-standards/iso-25010

Table 4. List of considered non-functional characteristics when maintaining workflows.

| characteristic | description |
| --- | --- |
| reliability | My workflow should operate failure-free for a specified period of time in a specified environment. |
| understandability | My workflow should be easy to read and understand. |
| testability | My workflow should be easy to test and debug. |
| security | My workflow should not have potential security weaknesses and actual vulnerabilities that could be (ab)used by an attacker. |
| modifiability | My workflow should be easy to change or adapt. |
| performance | My workflow should perform its functions within a specified amount of time and with efficient use of available resources. |
| reusability | My workflow or specific parts of it should be easily reusable in other contexts. |

The responses are summarised in Figure 5. The majority of respondents considered all seven characteristics to be at least *moderately* important, with a median of three characteristics being reported as *very important*. Four characteristics were reported as at least *moderately* important by a large majority of respondents: **reliability** (97.1%), **understandability** (87.1%), **security** (80.6%), and **testability** (80.9%). Notably, 80.2% of respondents considered **reliability** *very important*, and 63.7% did so for **security**. Even the three characteristics of "lesser" importance were generally valued by a majority of respondents, considering them at least *moderately* important : **modifiability** (73.5% of respondents), **performance** (71.6%), and **reusability** (53.7%).

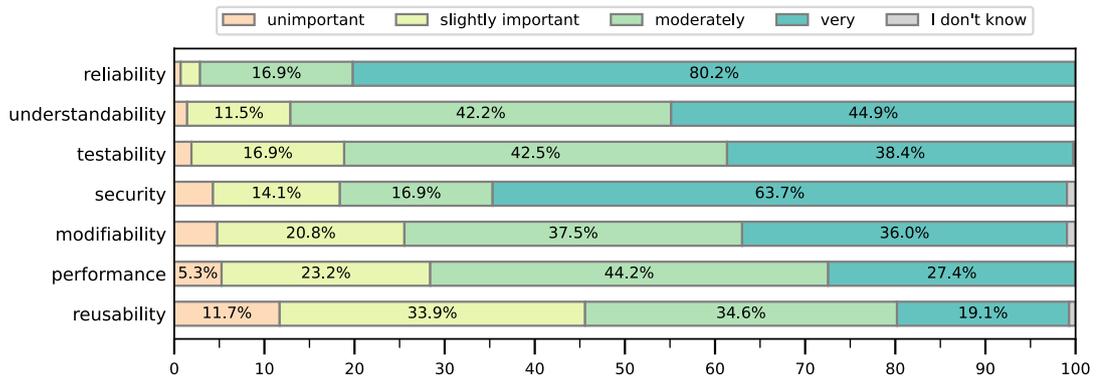

Fig. 5. Reported importance of non-functional characteristics when maintaining workflows.

We derived a partial order of the perceived *relative* importance of non-functional qualities when maintaining workflows, shown in Figure 6. It reveals a clear hierarchy, with **reliability** emerging as the paramount concern, underscoring the critical need for workflows to operate consistently and without failure within their designated environments. This is unsurprising, as unreliable workflows may lead to disruptions, wasted resources, and a loss of trust in the automated processes. The second most important non-functional quality is **security**, reflecting the growing awareness of potential vulnerabilities in automated pipelines and the significant risks associated with their exploitation. It is crucial to prevent workflows from having weaknesses that could be exploited by malicious actors. The third most important characteristic is **understandability**. As observed in prior studies [57, 69], GitHub Actions workflows present notable challenges due to the confusing and error-prone syntax of YAML files and strict indentation rules. The importance of understandability is further exacerbated by the difficulty in debugging workflow files, coupled with inadequate documentation and the lack of code completion and syntax highlighting.[14] Workflow maintainers frequently turn to Q&A forums such as Stack Overflow for assistance with GitHub Actions-related issues, and these questions have increased dramatically in recent years [69].

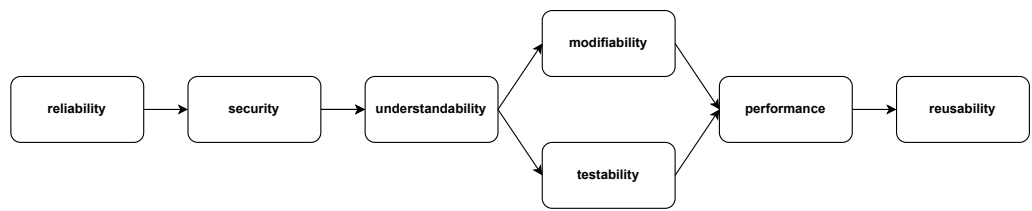

Fig. 6. Partial order of relative importance of non-functional characteristics when maintaining workflows, reflecting the relative choices made by respondents.

**Modifiability** and **testability** were perceived to be of lesser relative importance. Nevertheless, they are still considered at least *moderately* important by a large majority of respondents, reflecting the value of the ability to adapt and test workflows easily to ensure the correctness of modifications made to them. Indeed, multiple respondents complained about the difficulty of testing and debugging workflows due to the lack of proper tools. This observation that aligns with prior research [57].

**Reusability** and **performance** were deemed the least important of all characteristics, even though they were also considered at least *moderately* important by a majority of respondents. This suggests that workflow maintainers prioritise the other non-functional qualities over the potential for wider application of workflows or more efficient resource use, indicating a focus on immediate needs rather than proactive optimisation for future reuse or optimal performance. The low relative importance attributed to **performance** may be consequence of the availability of free GitHub-hosted runners for public repositories, making efficiency less of a concern unless resource limits are reached. The low relative importance of **reusability** suggests that GitHub Actions workflows do not yet embrace reuse to its fullest capacity. Workflow maintainers seem to underexploit GitHub Actions's reusability mechanisms, suggesting an untapped reusability potential, possibly due to unawareness and technical limitations in how these mechanisms are currently supported by GitHub.

Summary: For maintaining GitHub Actions workflows, respondents generally viewed all seven non-functional characteristics as important, and the need for reliable and secure operation were deemed most important. Understandability also holds significant importance, driven by the complexities of workflow configuration and the challenges in testing and debugging caused by inadequate tooling. Reusability and performance of workflows were considered to be of lesser relative importance.

### 4.4 Discussion: workflow automation practices

We observed a clear disparity in how different tasks are automated within GitHub Actions workflows. Core CI/CD activities like automated testing, compiling and building, code quality analysis, deployment, version and release management are widely adopted. Tasks such as internal reporting, infrastructure and configuration management, issue management, compliance checking, health and performance monitoring and external communication are automated considerably less frequently. This observed disparity in automation practices aligns with a study on the GitHub Marketplace, uncovering a misalignment between academic research that tends to focus on code quality and testing, while practitioners direct their attention toward more general CI/CD processes [56]. Similarly, Ghaleb et al. [26] observed a diversity in automation practices in Android projects, revealing that workflow configurations most frequently concentrate on initial build setup, with standard unit tests implemented in approximately half of the Android projects. Deployment automation was observed in only a minority of cases.

---

[14]Some of those limitations have been mitigated in recent versions of the GitHub UI.

Despite the importance of security analysis, our survey revealed that nearly 40% of all respondents do not use workflows to automate it. This aligns with findings from Angermeir et al. [5], who observed in a dataset of more than 8.4K OSS projects that only 6.8% implemented security activities within their CI pipelines. These observations are confirmed by a larger empirical study of CI/CD pipeline configurations in 28,770 OSS GitHub repositories [13]. Build and test activities are automated by the large majority of repositories (80%), while there is a low adoption of specialised practices such as security analysis (12%), containerisation (18%) and cloud deployment (4%). For GitHub Actions specifically, the use of workflows for security analysis amounts to 16%. This limited adoption of workflow automation for security analysis could be due to maintainers' potential unawareness of such mechanisms, a lower prioritisation of security, a conscious decision not to implement it, or the use of alternative tools that do not necessitate workflow automation, like GitHub's built-in CodeQL code scanner or third-party security scanners. The limited focus on security checks has notable implications for software quality and risk management. It delays vulnerability identification, going against the "shift left" principles [33] that advocate for early security integration to prevent more costly fixes later.

Taken together, the full potential of workflow automation for enhancing software quality, security, and operational efficiency has not yet been fully realised, negatively affecting product reliability and user experience. To broaden the scope of CI/CD automation, it is essential to raise awareness and provide targeted educational materials. Such initiatives should emphasise the benefits that can be achieved by automating a wider range of tasks.

> RECOMMENDATIONS:
>
> - Practitioners should harness the full capabilities of GitHub Actions by extending its use to automate repetitive tasks beyond the core CI/CD practices to cover more areas such as issue and PR management, internal reporting, health and performance monitoring, compliance checking, and external communication.
> - Researchers should analyse the reasons behind the limited automation of important tasks such as security analysis. Such studies could focus on understanding the barriers that prevent practitioners from automating them, including awareness, perceived complexity, or technical limitations.

Most respondents primarily create new workflows by either adapting their existing ones or by writing them completely from scratch. Reusing existing workflows created by others is also fairly common. In contrast, the adoption of tool-assisted workflow creation methods and starter templates is relatively low, suggesting that the discoverability, usability, or functionality of such automation aids should be improved. Despite the fact that starter templates are directly accessible through the Actions tab of one's repository in the GitHub UI, they are not used by a wide majority of the respondents. This adoption barrier could stem from the lack of tool support to customise starter templates to one's specific needs. As one respondent stated: *"We frequently use GitHub Actions for automating the creation and population of new repositories from templates. GitHub has templates, but they're not very powerful, so we have combined Python*

*Cookiecutter templates with GitHub Actions to automate the creation of new repositories to follow organisational standards."* Another limitation may be the lack of a sufficient number of starter templates to cover the diverse needs of workflow developers to automate more complex or more specialised tasks. At the time of writing, the GitHub UI listed only 174 starter templates across five categories: *Deployment* (27 templates), *Security* (79 templates), *Continuous Integration* (54 templates), *Automation* (5 templates), and *Pages* (9 templates).

> RECOMMENDATIONS: The low adoption of tool-assisted workflow generation or starter templates reveals a gap in the available tooling for facilitating workflow creation. This opens up improvement opportunities to provide a wider range of starter templates, increase their discoverability, and facilitate their parametrisation to project-specific contexts. It also calls for effective and commonly available automated workflow generators, e.g., based on AI or template generation tools.

For maintaining GitHub Actions workflows, respondents generally considered all seven non-functional characteristics as important. Reliability was considered the most crucial characteristic, directly reflecting its impact on the value and long-term sustainability of automated processes. This aligns with the critical role of workflows in automating software development. As one respondent highlighted: *"GitHub Actions is bad and generally rather unreliable. I rarely can reproduce builds consistently due to unexplainable errors in the depending Actions, surprise changes in the runner configuration and, most importantly, difficulty on addressing and debugging errors in the workflows."*

Additionally, testability and modifiability were recognised as important for ensuring the correctness of modifications and identifying potential issues. However, significant challenges with testability reveal a critical area where changes are greatly needed. The current reliance on trial-and-error underscores this inadequacy. Two respondents elaborated: *"Somehow the process of getting Actions running is always a trial and error. From experience I tend to tell to my team, that it costs 30 builds or releases to tune it to work as expected, from controlling modified paths, to running unit and integration tests, to validating coverage, etc. I think it would be good idea to mock the use of Actions from an IDE, so that it could be pre-validated before the launch, sometimes is just a small issue that a pre-validation or dry run on the client side, could avoid the launch in the Github side."* and *" In many cases I have to make blind changes in the workflow files hoping that they will fix the errors I observe. Additionally, figuring out whether there's something wrong with the runner is also hard, [...] Part of the problem is the fact that, as far as I know, there is no way to restart the workflow from a failed step.".*

The difficulties of testing and debugging GitHub Actions workflows not only reduce confidence but also hinder their overall effectiveness and maintainability, given their vital role in software development. For instance, two other respondents stated: *"My biggest issue with GitHub Actions and workflows is testing them. I know of https://github.com /nektos/act, but I have never managed to get it to work for anything but the most trivial workflows."* and *"It's an awful programming model that requires pushing junk commits and waiting for an undebuggable machine to spit some obscure error and try again and again."*

The analysis of automation tasks indicates a strong focus on core CI/CD activities. Large-scale resource usage studies revealed that the vast majority (over 91%) of computation resources in GitHub Actions workflows are consumed by testing and building activities [10]. This resource concentration highlights the need for optimising resource consumption for these activities. Bouzenia and Pradel [10] reported that effective optimisations, such as caching and setting timeouts, are substantially underused in practice, suggesting a considerable potential for improving workflow optimisation, testing, and debugging.

RECOMMENDATIONS:

- GitHub Actions needs improved testing and debugging tools. These should encompass capabilities such as simulating workflow executions with uncommitted changes, enabling restarts from failed steps, and providing the ability to inspect the runner's state in the event of a failure. Facilitating interactive exploration of the runner's state, including its variable configurations and logs at the point of failure would significantly ease debugging and accelerate the identification of root causes.
- GitHub Actions should enhance its real-time validation capabilities, currently available through the GitHub UI and IDE extensions. Such validation should go beyond mere syntax checking, e.g., by including comprehensive anaylisis of Action parameters to ensure that all mandatory inputs are provided and their values are properly formatted.

## 5  $G_2$: Understanding workflow reuse practices

Software reuse has become widely adopted in the last decade due to the prevalence of OSS libraries. Given that GitHub Actions workflow files can be regarded as "configuration as code", the second goal of this study is to understand the reuse practices followed by GitHub Actions workflow practitioners. Indeed, an important feature that has contributed to the popularity of GitHub Actions is the support it provides for avoiding duplication, by creating reusable automation components that can be shared and reused across workflows.[15] GitHub provides two main reuse mechanisms:

**Actions** are pre-written reusable components that can be run as part of a *step* in a *job* of the workflow. They are versioned, and must be invoked through the `uses:` keyword. The use of such components prevents workflow maintainers from having to write the commands by themselves (through the `run:` keyword). These Actions come in three flavours: (i) *JavaScript Actions* are developed in JavaScript that can be distributed through the GitHub Marketplace and used directly from any GitHub repository providing such an Action; (ii) *Docker container Actions* aim to run Docker container images that can be stored in a GitHub repository or some container registry; (iii) *Composite Actions* are indistinguishable in usage from *Docker container Actions* or *JavaScript Actions*. They represent a special type of Action designed to group multiple steps together, encapsulating a series of commands or dependencies on other Actions. They are built directly using the YAML-based workflow syntax, just like an ordinary workflow. This enables sharing a sequence of workflow job steps by allowing them to be run as a single step across various workflows.

**Reusable workflows** allow calling and running a complete workflow from within other workflows to avoid copy-pasting code from one workflow to another. They are invoked using the on: workflow_call: keywords.

Beyond these reuse mechanisms, we also specifically asked respondents about their *copy-paste* coding practice within and across workflows, given that it remains a prevalent way of reusing of code in software development practice [34]. This practice is informal and often inefficient, possibly leading to increased maintenance burden, as well as the propagation of bugs and security vulnerabilities, due to the absence of tools to systematically manage copy-paste reuse [6, 34, 40]. Understanding the reasons behind or against copy-pasting will offer valuable insights into developer preferences for actual reuse practices.

---

[15]See https://docs.github.com/en/actions/sharing-automations and https://docs.github.com/en/actions/sharing-automations/avoiding-duplication.

## 5.1 Origins of workflow reuse

We started by asking the respondents how frequently they resort to copy-paste reuse and the built-in reuse mechanisms of GitHub Actions. Table 5 lists the different options proposed to the respondents. Separating between custom Actions and reusable workflows, we focused on the "origin" of these reusable components, distinguishing between those that were created by the respondents themselves and those that were created by someone else.

Table 5. List of origins of reuse in GitHub Actions workflows.

| reuse mechanism | description |
|---|---|
| own Action | I use an **Action** that I have created myself. |
| other's Action | I use an **Action** developed by someone else. |
| own reusable workflow | I use a **reusable workflow** that I have created myself. |
| other's reusable workflow | I use a **reusable workflow** that was developed by someone else. |
| copy own workflow | I **copy and paste** parts of workflows from my own repository. |
| copy other's workflow | I **copy and paste** parts of workflows from someone else's repository. |
| copy other source | I **copy and paste** parts from some other source (e.g., forum, Q&A platform). |

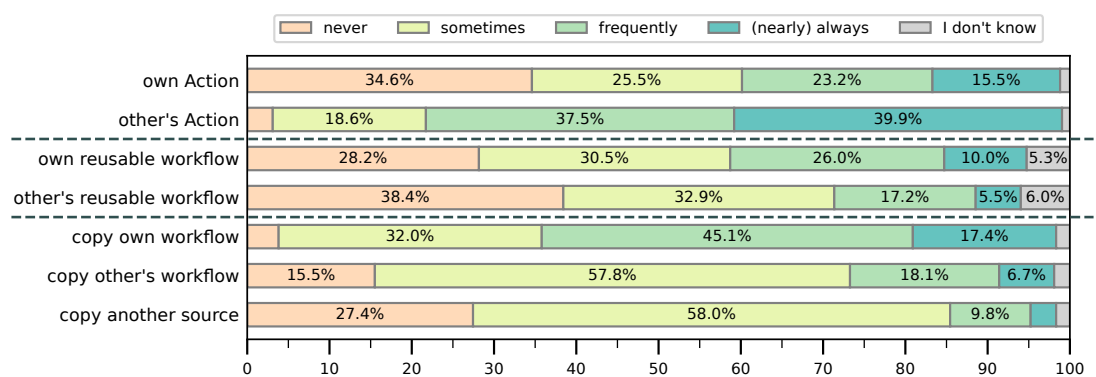

Fig. 7. Reported frequency of using workflow reuse mechanisms.

Respondents reported a median of six out of seven different reuse mechanisms used when creating or maintaining workflows. Of these, a median of three mechanisms were reported to be used *frequently* or *nearly always*. Figure 7 summarises the responses. A clear pattern emerges in how these mechanisms are employed. Reusing someone **other's Action** is the most frequent, with 77.4% respondents reporting it at least *frequently*. This aligns with earlier quantitative empirical evidence showing that a significant proportion of workflow job steps (51.1%) use an Action, with almost half (48.3%) being third-party Actions [21]. When comparing the reuse of one's **own Action** with reusing someone else's Action, we observed a significant disparity: 64.2% respondents relied on their own Actions at least *sometimes*, while the overwhelming majority of respondents (95.4%) using others' Actions at least *sometimes*. This can be explained by the quantitative observation that nearly all workflows rely on one of GitHub's provided Actions, such as the almost indispensable actions/checkout [21].[16] The high proportion of respondents who *never* rely on their own Actions can be explained by the difficulty in developing Actions. Based on a survey with 44 Action developers, Saroar and Nayebi observed that 75% of respondents faced one or more challenges in developing, testing, or debugging Actions due to limited documentation and support and the poor design of the Actions framework [57].

---

[16]https://github.com/marketplace/actions/checkout

The data reveals a general tendency to avoid reusable workflows. 28.2% of respondents reported that they *never* use their **own reusable workflows**, while 38.4% *never* use **others' reusable workflows**. Reusable workflows which enable the invocation and execution of complete automation pipelines from other workflows, are less frequently employed compared to reusing Actions or copy-pasting. More precisely, only 36% of respondents reported using their own reusable workflows at least *frequently*. Reliance on others' reusable workflows was even less common, with merely 22.7% doing so at least *frequently*. The reasons behind the adoption or avoidance of reusable workflows are discussed in detail in the next section (Section 5.2). For instance, one respondent articulated a key technical limitation: *"A reusable workflow has to be an entire job; it cannot just be a step. This limits its usefulness. But even then, I should probably use it more often."*

Copy-pasting remains a prevalent practice. The second most prominent mechanism is **copy own workflow**: 62.5% of all respondents at least *frequently* copy parts from one of their own workflows. **Copy other workflow** and **copy another source** occurs less commonly. A frequent itemset analysis revealed that 81% of respondents at least once employed a combination including reusing **other's Action**, **copy own workflow**, and **copy other's workflow**. This underscores the common practice of integrating multiple reuse mechanisms.

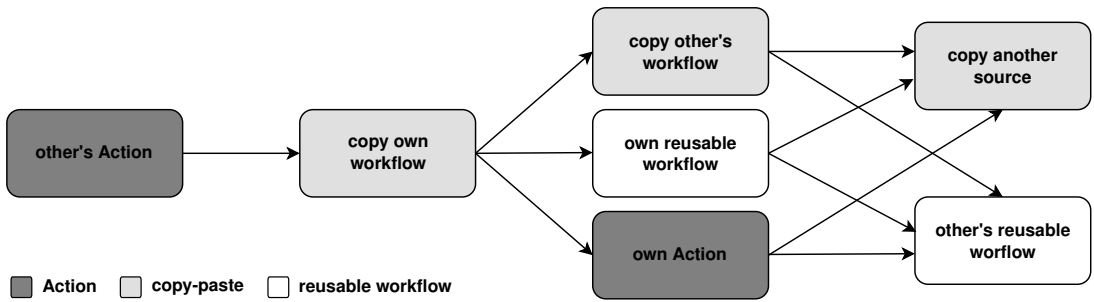

Fig. 8. Partial order of frequent reuse mechanisms, reflecting the relative choices made by respondents. The shading reflects the three groups of reuse mechanisms presented in Table 5.

We derived a partial order of respondents' relative frequency of workflow reuse mechanisms. Figure 8 confirms the clear preference to **copy own workflow** as opposed to **copy other's workflow** or **copy another source**. This preference might indicate a greater comfort or trust level with previous self-written and understood code. It does not imply that relying on other reuse mechanisms is uncommon, since each of the seven workflow reuse mechanisms reported in Figure 7 is reported to be used at least *sometimes* by a majority of respondents.

Two respondents signaled using a workflow reuse mechanism that is currently not part of GitHub itself, namely the auto-generation of workflows: *"I use reusable Kotlin snippets [...for generating GitHub Actions workflows]"* and *"Our workflows are often auto generated by a separate tool which can abstract over different CI services, and is simplified because it only supports specific tasks that we need. As a result our actual GitHub workflows don't re-use as much as they could, and have a lot of "copy paste" but it's just auto generated."* One respondent also reported the use of generative AI: *"CoPilot suggested it, I approve it, then I copy/paste it."* One can reasonably expect considerably more usage of generative AI in the future, given the increasing availability of such tools for supporting collaborative software development.

SUMMARY: Actions are the most frequent reuse mechanism, particularly Actions sourced from others, demonstrating a strong reliance on external third-party components. Creating and using one's own Actions is less common. Reusable workflows remain underused, whether they be self-created or sourced from others. Copy-pasting remains a prevalent mechanism, especially from one's own code. Developers frequently combine different reuse mechanisms, indicating a pragmatic approach. Looking ahead, auto-generated workflows and generative AI are emerging as new avenues for reuse.

### 5.2 Motivations for copy-pasting

GitHub offers varying levels of reusability to accommodate different automation needs, allowing practitioners to select the most appropriate level of abstraction for developing and maintaining their workflows. *Actions* provide granular reusability to encapsulate individual tasks or closely related commands, running either JavaScript code or deploying Docker containers for specific environments. *Composite Actions* enable the reuse of an entire sequence of workflow job steps as a single logical unit. Finally, workflows can rely on *reusable workflows* to invoke and execute complete automation pipelines. Despite these three reuse mechanisms, practitioners might still resort to copy-pasting practices. To understand if and why they do so, we asked three distinct questions, each one focusing on a specific scenario and a corresponding level of reusability:

(1) *Do you tend to copy and paste* **commands in a** `run:` **step** *instead of using an Action? If yes, why?*
    This question aimed to understand why practitioners might copy and paste inline commands rather than using encapsulated *Actions* for specific tasks.

(2) *Do you tend to copy and paste* **multiple steps** *instead of using a composite Action? If yes, why?*
    Here, we sought to determine why practitioners might copy and paste sequences of steps in workflow jobs rather than abstracting them into a reusable *composite Action*.

(3) *Do you tend to copy and paste a* **job** *instead of using a reusable workflow? If yes, why?*
    This final question explored the reasons behind copy-pasting entire jobs rather than leveraging the highest level of reusability offered by *reusable workflows*.

We presented a predefined list of reasons for preferring copy-paste and for not adopting the considered reuse mechanism, obtained through the process explained in Section 3.1. Table 6 lists the reasons presented to respondents regarding their preference for copy and pasting over the three built-in reuse mechanisms, as well as the reasons for not adopting these mechanisms. Respondents were presented checkboxes to select whether they agreed with each reason.

For each of the three reuse mechanisms, Figure 9 provides a radar plot summarising the received responses. Since the main aim is to understand why a particular reuse mechanism is underused, we restricted the analysis to only those respondents that reported relying on copy-paste instead of the considered reuse mechanism, i.e., we excluded all respondents that reported not to rely on copy and paste. For the *Actions* mechanism, the considered subpopulation corresponds to 78.0% of all respondents, for composite Actions this was 79.7%, and for reusable workflows 74.9%. Below we provide the interpretation of each radar plot.

*Actions.* 59.6% of the respondents that tended to copy commands directly into `run:` steps rather than utilising *Actions* found this practice more *convenient*, and 73.3% believed it offered greater *control and flexibility*. One respondent stated *"It is unclear to me when and how [using Actions] would be more convenient than a shell script."*. Despite Actions being a common reuse practice, several factors hinder their wider adoption when compared to the preference for copy-pasting commands in a step. The biggest reported challenge was *undiscoverability* of suitable Actions (52.8%). Even with the GitHub Marketplace available, many practitioners struggle to locate existing Actions that genuinely meet their requirements. A notable proportion of respondents (38.2%) also considered that using an Action for very simple tasks introduced unnecessary *complexity*. While very few respondents were completely unaware (2.5%) or unfamiliar (12.4%) with Actions, these knowledge gaps still play a role. Finally, a *lack of trust* was not a major issue, though 10.6% of respondents were hesitant to rely on Actions.

Table 6. List of reasons for copy-pasting instead of using GitHub Actions reuse mechanisms.

| reason for copy-pasting | description |
| --- | --- |
| no copy-paste | I do not copy and paste. |
| convenience | Copy pasting is more convenient (e.g., easier, faster). |
| control and flexibility | Copy pasting gives me more control and/or flexibility. |

| reason for not adopting reuse mechanism | description |
| --- | --- |
| unawareness | I am unaware of the mentioned reuse mechanism. |
| unfamiliarity | I am not sufficiently familiar with the mentioned reuse mechanism. |
| undiscoverability | I cannot find a suitable artefact (i.e., Action, composite Action or workflow) to reuse. |
| complexity | Adopting the mentioned reuse mechanism induces additional complexity. |
| lack of trust | I do not trust the mentioned reuse mechanism. |

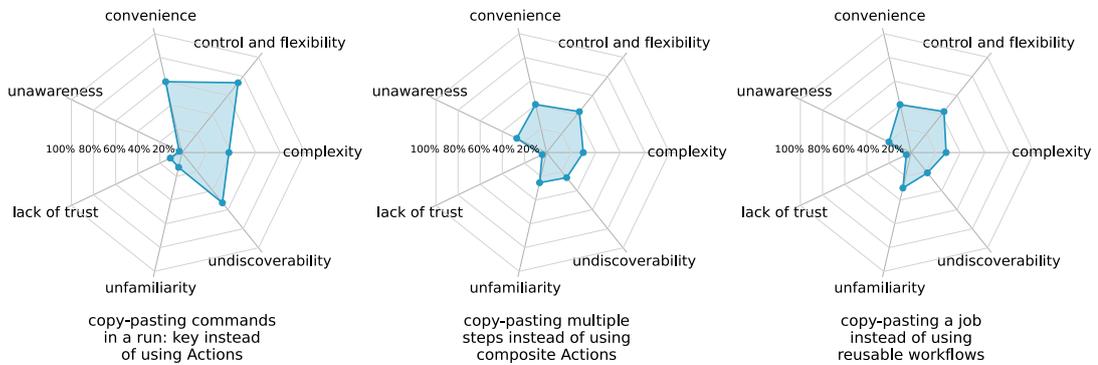

Fig. 9. Reported reasons for copy-pasting instead of using an available reuse mechanism. From left to right, the radar plots represent *Actions*, *composite Actions*, and *reusable workflows*, respectively. Respondents that reported not to rely on copy and paste were excluded from the analysis.

**Composite Actions.** 40.5% of the respondents that tended to copy and paste sequences of steps rather than creating or using a *composite Action* found this practice more *convenient*, and 42.9% reported it offered more *control and flexibility*. Copying a block of steps in the workflow file can feel faster and offers more immediate oversight than setting up and referencing a separate composite Action. Many hurdles stand in the way of using composite Actions more broadly. 26.9% of respondents were *unaware* of composite Actions, and 25.4% were *unfamiliar* with them. The perceived *complexity* (29.9%) also plays a part; practitioners might feel that the effort involved in creating and maintaining a composite Action outweighs the benefits. Furthermore, *undiscoverability* of suitable composite Actions (26.3%) can be challenging. *Lack of trust* was of little concern (3.9%).

***Reusable workflows.*** 40.3% of the respondents that tended to copy and paste entire jobs instead of leveraging reusable workflows did so out of *convenience*, and 43.0% because it offered greater *control and flexibility*. They believe that duplicating a job definition is simpler and quicker than setting up and calling a reusable workflow. The reasons why largely stem from *unawareness* of the reuse mechanism (20.0%) or *unfamiliarity* with it (29.8%). The perceived *complexity* (28.5%) of reusable workflows also discourages their use, as managing inputs, outputs, and the overall call structure can seem more involved than simply copying the job definition within the same or another workflow file. *Undiscoverability* of suitable reusable workflows presented a deterrent for 21.3% of these respondents. *Lack of trust* was of little concern (4.3%).

Below, we delve deeper into the reasons listed in Table 6 for not adopting a reuse mechanism and relate this to some findings of prior research (e.g., [57]). This time, we focus on the entire population or respondents since the goal is to compare these reasons regardless of whether respondents have used a particular reuse mechanism or not. As a consequence, the percentages reported below differ from the percentages in the previous analysis.

***Unawareness and unfamiliarity.*** Almost no respondents (2.6%) were *unaware* of the **Actions** reuse mechanism. For the other two reuse mechanisms the proportion of unawareness was considerably higher: 21.7% for **composite Actions** and 15.0% for **reusable workflows**. Only a small proportion of respondents were *unfamiliar* (10.3%) with the use of **Actions**. Again, this proportion was higher for the reuse mechanism of **composite Actions** (20.8% of unfamiliar respondents) and **reusable workflows** (22.7%). We cross-linked these results with the respondent's self-reported familiarity with GitHub workflows (see Table 1), and observed that **Actions** are a well-established reuse mechanism: only 16.2% of respondents that were *moderately familiar* with GitHub workflows were *unaware* or *unfamiliar* with **Actions**. For respondents that were *very familiar* with GitHub Actions workflows, this was 8.8%. The mechanism of **reusable workflows** was considerably less known: 44.9% of respondents that were *moderately familiar* with GitHub Actions and 29.6% that were *very familiar* reported being *unaware* of or *unfamiliar* with this reuse mechanism. **Composite Actions** were even less known to respondents: 51.5% of *moderately familiar* and 31.4% of *very familiar* respondents reported being *unaware* or *unfamiliar* with this reuse mechanism. This raises the need for increasing awareness of both underused mechanisms.

***Undiscoverability.*** More than half of all respondents (52.0%) considered *undiscoverability* as a problem for at least one of the three reuse mechanisms. A significant proportion of respondents indicated that they could not easily find an appropriate **Action** to reuse (45.6%), with similar issues reported to a lesser extent for **composite Actions** (22.2%) and **reusable workflows** (16.5%). The challenge of discovering suitable Actions confirms earlier observations from a survey of Action developers [57], where nearly 57% stated creating new Actions because they could not find existing ones meeting their requirements. Furthermore, a quarter of these developers believed their needs to be unique and that existing Actions offered insufficient functionality.

***Complexity.*** 43.7% of all respondents reported that adopting a reuse mechanism would introduce additional *complexity*. For Actions specifically this was 31.0%. As some respondents elaborated: *"If an Action is actually really simple, it's better to copy it, rather than depend on an external Action"* and *"The commands I run are either too simple for an Action (cargo build) or too complicated to re-use."* One respondent even related *complexity* to the non-functional characteristic of *testability* explored in Section 4.3: *"Github Actions aren't a typed language, it's YAML config. You want to minimize the complexity there because you cannot test it without running it. The more complex, the more edge cases that could bite you."* This inclination to avoid reuse mechanisms aligns with observations from a prior qualitative study [57] that a subset of workflow maintainers intentionally avoided using Actions when the perceived cost or effort outweighed the benefits, when it increased *complexity*, or due to a preference for manual execution of certain tasks.

Dependency problems were also perceived to be a major source of *complexity*, making respondents decide not to reuse Actions in their workflows. This was reflected through many responses provided in the optional free-text field. Several respondents complained about problems they encountered due to the dependency hell incurred by the need to manage all these additional (sometimes even indirect) dependencies. One respondent argued: *"Dependencies are often a liability."* Another one explained that *"For simple tasks, I feel it is better to avoid an additional dependency on an external Action."* Similarly, some respondent complained about *"Yet another dependency to manage"* and another one *"Indirection makes it harder to reason about"*. And a slightly longer answer *"Most steps are less than 5 lines of bash. I don't want complexity spread across multiple locations, unless it's an official Action (feels like a standard library), unless it's actually not worth maintaining the complexity myself (similar to the decision to include dependencies in software itself), and even then I'd prefer to use something well established."*

**Lack of trust.** did not appear to be a primary deterrent to adopting reuse mechanisms. Only a small percentage of all respondents (10.7%) confirmed that lack of trust was a reason for not adopting at least one of the reuse mechanisms. For **Actions** specifically this was 9.5%, for **composite Actions** 3.6%, and for **reusable workflows** 3.1%. One respondent told *"I do not trust the authors of Actions."* Another one *"I mainly trust MY actions, not others"* and a third one *"There are only a handful [of Actions created by others] I trust enough to use."*

The fact that the overwhelming majority of respondents do trust the workflow reuse mechanisms seems somewhat counter-intuitive, given that Figure 6 revealed security to be the second most important characteristic considered when maintaining workflows. This is confirmed by one respondent warning about the *"supply chain risks of externally-maintained Actions"*. Previous empirical studies have indeed confirmed that trusting and relying on external, third-party Actions could introduce security vulnerabilities [7, 22, 52]. *dawidd6/action-download-artifact* serves as an example of a compromised **Action** with a high-severity artefact poisoning vulnerability, as indicated by GitHub advisory reports.[17] Concerns regarding security also extend to **composite Actions** and **reusable workflows**. For instance, *tj-actions/branch-names* is a **composite Action** that has been identified as critically vulnerable to a command injection attack according to the GitHub advisory database.[18] Similarly, *kartverket/github-workflows* is a **reusable workflow** with a high-severity vulnerability reported in the GitHub advisory database,[19] affecting all users of its *run-terraform* **reusable workflow**. A malicious actor could potentially submit a PR with a malicious payload, leading to the execution of arbitrary JavaScript code within the workflow's context. Furthermore, the **reusable workflow** *canonical/get-workflow-version-action*[20] has a high-severity vulnerability, as per the GitHub advisory database,[21] which can leak a partial GitHub Token in exception output. These examples serve to illustrate that important security vulnerabilities do occur with GitHub's reuse mechanisms, highlighting a potential disconnect between practitioners' trust and the demonstrable security risks associated with relying on external components, underscoring the importance of careful scrutiny and awareness when adopting reuse mechanisms.

---

[17]https://github.com/marketplace/actions/download-workflow-artifact
[18]https://github.com/advisories/GHSA-8v8w-v8xg-79rf
[19]https://github.com/advisories/GHSA-f9qj-7gh3-mhj4
[20]https://github.com/canonical/get-workflow-version-action
[21]https://github.com/advisories/GHSA-26wh-cc3r-w6pj

Summary: The majority of respondents copy and paste code rather than using GitHub Actions' built-in reuse mechanisms. Convenience, control, and flexibility are indicated as the main reasons for this preference. More than two out of five also feel that adopting these reuse mechanisms adds unnecessary complexity, with dependency issues being repeatedly cited as a major source of complexity. More than half of all respondents struggled with finding suitable reusable artefacts. Awareness of and familiarity with composite Actions and reusable workflows are considerably lower than for Actions, even among experienced users. Lack of trust was not identified as a primary reason for avoiding reuse mechanisms, despite security being considered an important concern and the fact that known security vulnerabilities have been documented for all three types of reusable components.

### 5.3 Important characteristics for selecting and using Actions

Workflows commonly employ reusable Actions [21], which serve as reusable components for task automation. Section 5.1 and Section 5.2 highlighted two key trends in Action usage. Firstly, practitioners tend to use *Actions developed by others* more frequently than their own. Secondly, *discoverability* remains a significant challenge, making it difficult to locate the most suitable Action. These findings underscore the need for effective Action selection, suggesting that specific characteristics, beyond mere functional adequacy, play a crucial role. Selecting appropriate Actions presents a non-trivial challenge given the number of Actions on the GitHub Marketplace, the heterogeneity in non-functional characteristics, and the absence of tools for assessing the functional and non-functional qualities of Actions. To gain deeper insight into the criteria that guide workflow maintainers in selecting and using Actions, we asked respondents which non-functional characteristics they consider important for this purpose. Drawing upon prior investigations into the adoption of reusable OSS components [4, 42, 44, 48], we identified a set of nine key characteristics in Table 7.

Table 7. List of considered non-functional characteristics when selecting and using a reusable Action in a workflow.

| characteristic | description |
| --- | --- |
| reliability | The failure-free operation of the Action for a specified period of time in a specified environment. |
| documentation | The clarity and consistency of the Action's documentation, such as installation and usage instructions. |
| maintenance | Evidence that the Action is modified to correct faults, improve performance or other attributes, or adapt to a changed environment. |
| security | The Action should not have security weaknesses and vulnerabilities that could be (ab)used by an attacker. |
| performance | The ability of the Action to perform its functions within a specified time and with efficient use of resources. |
| quality | The overall quality of the Actions code base (such as its understandability, simplicity, modularity) and the availability of tests. |
| popularity | Indicators that signal that there is a community interested in the Action (e.g., as reflected by a high number of stars, watchers, or forks). |
| responsiveness | The extent to which the Action's maintainers actively react to user requests or engage in discussions. |
| license compatibility | The compatibility of the Action's license with the intended use of the Action. |

Respondents were asked to respond to the following question: *How important do you consider the following characteristics when using an Action in your workflow?* The results, shown in Figure 10, indicate that all nine characteristics were considered at least *slightly* important by a majority of respondents (the first seven characteristics being even considered at least *moderately* important by them). Respondents found a median of eight characteristics to be important, of which a median of three characteristics were considered *very important*.

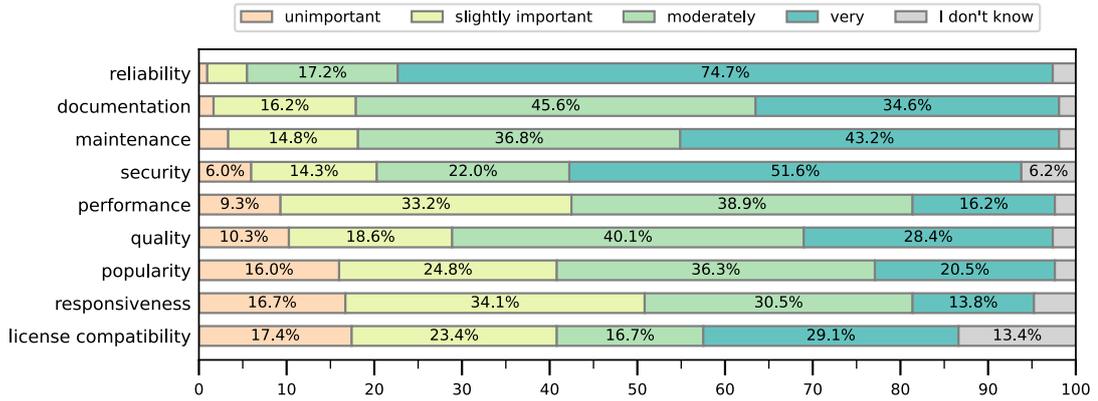

Fig. 10. Importance of considered characteristics when selecting and using a reusable Action in a workflow.

Four characteristics were considered at least *moderately* important by a large majority of respondents: 91.9% for **reliability**, 80.2% for **documentation**, 80.0% for **maintenance**, and 73.6% for **security**. Two of these characteristics were even considered as *very important* by a majority of respondents: **reliability** (with 74.7%) and **security** (with 51.6%). One respondent reflected on the **maintenance** aspect as follows: *"I like the use of reusable Actions, but I also don't enjoy touching the CI scripts frequently, so it's highly important to me to be on a version that remains compatible and just works for as long as possible. In other words, low maintenance of my CI scripts has a very high priority to me."*

Only two characteristics were considered at least *moderately* important by less than half of the respondents: **license compatibility** (45.8%) and **responsiveness** (44.3%). For the former, 13.8% of respondents reported not knowing whether license compatibility was of any importance. A respondent even detailed: *"What on earth does license compatibility mean for an action? License compatibility with what? [...] I've never seen an action that would be licensed in a way that would prevent you from using it in any workflow, in any repository. Do these actually exist?"* Indeed, while license compatibility is a well-understood concern for software components generally, its meaning in the context of GitHub Actions far is much less evident. This may explain the high standard deviation among respondents, and many respondents expressing uncertainty or finding it unimportant. We consulted three academic licensing experts about the possible implications of an Action's license preventing its use in workflows, without clear conclusions. One perspective is that merely having a configuration file for an Action might not inherently grant explicit permission for use. The converse perspective argues that simply executing Actions might generally be permissible even without a specific license, as users typically let GitHub execute them on their behalf rather than obtaining them directly. The GitHub Terms of Service grants a nonexclusive, worldwide license to use publicly available content on GitHub. In short, the translation of traditional software licensing concerns to the unique operational model of GitHub Actions remains an area of ambiguity.

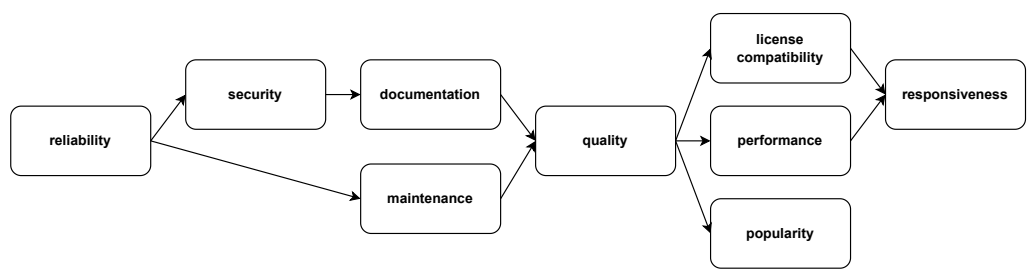

Fig. 11. Partial order of important characteristics for selecting and using Actions

We established a partial order of respondents' relative importance of the selected non-functional characteristics. As shown in Figure 11 the characteristic that was perceived as most important is **reliability**, reflecting practitioners' emphasis on using Actions that exhibit failure-free operation. Directly following **reliability** are three characteristics comprising **security**, **maintenance**, and **documentation**. The next important characteristic is **quality**. The remaining part of the partial order includes characteristics that remain relevant but are perceived as less important than the preceding ones when selecting Actions for workflows: **performance**, **popularity**, **license compatibility**, and finally **responsiveness**. Our findings confirm observations from prior work [57] that practitioners favour: bug-free Actions, a preference related to **reliability**; Actions from verified creators, highlighting the importance of **security**; properly documented Actions; and Actions with higher star ratings and more contributors, reflecting their **popularity**.

SUMMARY: All nine non-functional characteristics for selecting and using Actions were valued, with clear priorities emerging. Reliability was paramount, with nearly three-quarters of respondents finding it very important, highlighting a strong desire for failure-free Actions. Security, maintenance, and documentation followed, consistently rated as at least moderately important by most. These top four characteristics reflect core priorities: stable, secure, and well-supported Actions that minimise ongoing effort. Performance, popularity, and responsiveness were generally perceived as of less relative importance. It remains unclear what license compatibility actually means in the context of GitHub Actions.

### 5.4   Issues with using Actions

Reusing Actions in workflows is a common practice, allowing workflow maintainers to leverage existing functionality and avoid reinventing the wheel [21]. On the other hand, reusing Actions can introduce various issues such as outdated or deprecated Actions [20], breaking changes, and security vulnerabilities [52]. We therefore aim to gain a deeper understanding of how frequently workflow maintainers face such issues when using Actions in their workflows. We identified the main issues workflow maintainers could face when using Actions in their workflows. These issues, listed in Table 8 are heavily inspired by a comprehensive catalogue of challenges that can arise when depending on third-party open-source packages or libraries [47]. We asked respondents to report how frequently they have been facing these issues, and whether they have encountered any other issues.

Figure 12 reveals that, generally speaking, respondents are not facing issues too frequently, with *never* and *sometimes* being the predominant responses for all issues. Nevertheless, an overwhelming majority of respondents (96.7%) had faced at least one of these issues in their workflows. Within this group, a significant proportion (41.0%) reported encountering at least one of the issues *frequently* or *nearly always*. Furthermore, a majority of respondents (56.3%) indicated facing at least four of these issues at least *sometimes*. This not only demonstrates that almost all respondents have encountered an issue, but also that many of them experienced multiple issues when relying on Actions.

Table 8. List of encountered issues when using reusable Actions in workflows.

| issue | description |
|---|---|
| outdated | My workflow is not using a recent version of an Action. |
| deprecated | My workflow is using a deprecated Action (version). |
| breaking | My workflow stops working because of a breaking change. |
| unreliable | My workflow is no longer working like it used to. |
| unmaintained | My workflow is using an unmaintained or abandoned Action. |
| vulnerable | My workflow has become vulnerable. |
| license issue | My workflow is using an Action with an incompatible license. |
| unavailable | My workflow stops working because the Action is no longer available. |

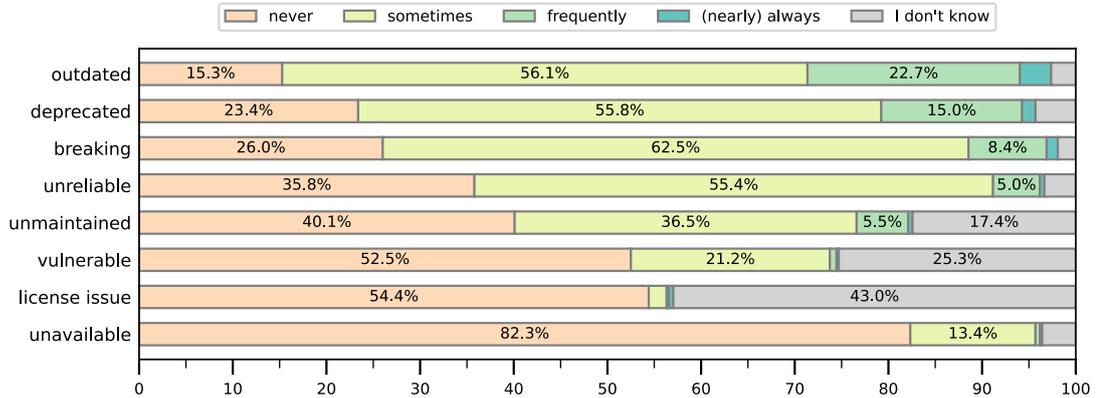

Fig. 12. Frequency of encountered issues when using a reusable Action in a workflow.

The most frequently reported issues are related to Action **outdatedness**, **deprecation** and **breaking changes**. Most respondents (56.1%) reported experiencing **outdated** Actions *sometimes*, while over a quarter (26.0%) encountered them *frequently* or *nearly always*. One respondent explained **not** to rely on Actions in workflows because *"it is simpler for basic commands, and does not go out of date"*. Another respondent complained that *"some actions are version locked"*. **Breaking** Actions were experienced *sometimes* by a majority of respondents (62.5%), while 9.5% reported encountering them *frequently* or *nearly always*. One respondent stated that *"some third party actions are just broken"*. Another one expressed the concern that *"Actions created by others are, generally, terribly designed and will almost certainly break in future versions."* Yet another one said *"I think breaking changes can be the most typical problem."* The issue of **deprecated** Actions or Action versions was reported *sometimes* by most respondents (55.8%), and a further 16.5% reported encountering them *frequently* or *nearly always*. Most respondents (55.4%) reported having *sometimes* suffered from the **unreliability** of the Actions they are using. Issues regarding **unmaintained** or abandoned Actions were experienced *sometimes* by 36.5% of respondents. For example, one respondent stated that *"relevant Actions are unsupported"*. Only 22.2% of respondents reported having at least *sometimes* faced an issue with their workflows having become **vulnerable** by using an Action, and 25.3% were not aware if such an issue arose or not. The issue of Actions becoming **unavailable** was uncommon, as it was encountered at least *sometimes* by only 14.1% respondents. **Licensing issues** were extremely uncommon, with only 2.6% of respondents having experienced this problem at least *sometimes*, and 43.0% being unaware of such issues.

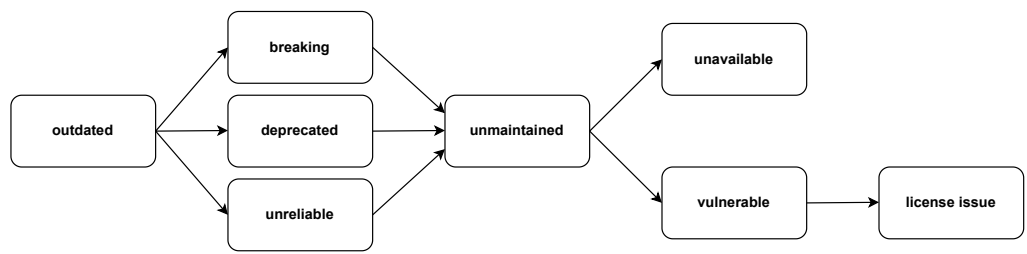

Fig. 13. Partial order of frequency of encountered issues when using a reusable Action in a workflow.

We established a partial order of the relative frequency of issues encountered by respondents when using Actions in workflows. This partial order, presented in Figure 13, suggests that issues with Actions in workflows are not equally distributed and some are more frequent than others. For instance, the issue of Actions being **outdated** has a higher relative frequency than those that are **deprecated** or that introduce **breaking** and **unreliable** workflows. These, in turn, are more frequent relative to the issue of Actions being **unmaintained**.

The reported frequencies of practical challenges in effectively using reusable Actions in workflows resonate strongly with the difficulties highlighted by a prior qualitative study on workflow automation [57]. That study identified key difficulties in testing and debugging automated workflows, the complexity of Actions for certain tasks, and inadequate documentation. These problems can be seen as contributing factors, and in some cases even root causes, for the specific issues faced by respondents. For instance, the complexity of Actions and a lack of clear documentation could make it harder for users to diagnose and resolve issues with reusable Actions.

SUMMARY: Nearly all respondents reported encountering issues when using Actions in workflows, with 2 out of 5 respondents facing at least one issue frequently. The most frequent issue was outdated Actions, but deprecated Actions, breaking changes and unreliable workflows were also reported by over half of the respondents at least sometimes. Issues related to unmaintained (or abandoned) Actions and vulnerable workflows were much less common. The large majority of respondents did not face, or were unaware of licensing issues and unavailability of Actions.

## 5.5 Discussion: workflow reuse practices

Driven by the prevalence of massive libraries of OSS components, the last decades have seen widespread adoption of software reuse [19]. GitHub also supports this practice for workflows by enabling the use of reusable Actions and reusable workflows. Our study into GitHub's reuse mechanisms, including the common practice of copy-pasting code within and across workflows, revealed several insights. Actions emerged as the most frequently used mechanism, particularly those sourced from third-party providers. Conversely, creating and using one's own Actions is less common, and reusable workflows are even less used. Reusable Actions provide a modular approach to composing and automating

GitHub Actions workflows. However, configuration complexity remains a significant challenge for workflow maintainers. Recent studies have found that GitHub Actions workflow files are often complex and lack consistent conventions, which can make it difficult for maintainers to effectively adopt and reuse existing mechanisms [26].

A majority of the respondents admitted copy-pasting code rather than utilising GitHub Actions' built-in reuse features. This preference was primarily driven by perceived convenience, control, and flexibility. The less frequent adoption of reusable workflows and composite Actions, even among experienced respondents, suggests that they are either unaware of these two mechanisms or face technical limitations in using them. Many respondents reported that adopting the built-in reuse mechanisms introduces unnecessary complexity, suggesting a need for more convenient-to-adopt reusable components. Furthermore, developers often struggle to find suitable reusable artefacts. As indicated in a previous study [57], the majority of Actions are simply made available in a repository and never published on the Marketplace, making them hard to discover. The challenges of discovering and selecting suitable components are being addressed by recent advances in automated tooling. For instance, Nguyen et al. [49] demonstrated the feasibility of using machine learning to automatically categorise Actions based on their README files, addressing the issue of unclear purpose and poor categorisation of Actions. Building on this improved clarity, Huang and Lin [32] introduced CIGAR, a contrastive learning model designed to accurately recommend relevant Actions, significantly improving the selection process compared to the default Marketplace experience.

Despite its convenience, the practice of copy-pasting can lead to maintenance issues [35, 53] such as *bug propagation*, where bugs in copied code may propagate to all locations where the code has been pasted. Code cloning also hinders refactoring, thereby lowering the code quality and increasing its maintenance effort. Nevertheless, previous studies have also pointed out possible advantages of code cloning, such as reduced development time and increased control and flexibility [35, 53]. Code cloning may also represent the only viable option for adding or enhancing a functionality in languages lacking robust reuse and abstraction mechanisms. How copy-pasting affects GitHub Actions workflows specifically in the short and long term remains to be investigated.

RECOMMENDATIONS:

- GitHub should enhance the discoverability of reusable components. This could involve improving the Marketplace's search and filtering capabilities, or providing curated lists of recommended components for both Actions and reusable workflows.
- GitHub should simplify the creation and integration of reusable components, to increase the adoption of the built-in reuse mechanisms. This can be achieved by providing comprehensive documentation, more examples and a set of best practices to lower the complexity.
- GitHub should promote its less popular reuse mechanisms of reusable workflows and composite Actions. This could involve providing more educational resources and practical examples, highlighting their advantages.
- Researchers should study code cloning practices in the context of GitHub Actions workflows, in order to understand its positive or negative impact on workflow quality and maintainability in the short and long term.

GitHub Actions workflows frequently rely on reusable Actions [20, 21]. Assessing the suitability of an Action for reuse presents a significant challenge, primarily due to a lack of tools and metrics for evaluating their non-functional qualities. Such tools would provide workflow developers with a more objective basis when looking for Actions to reuse in their workflows, potentially leading to more informed decisions and greater trust.

In general, reusable software components enable practitioners to leverage existing functionality. However, they also tend to introduce a wide range of dependency issues such as outdated or deprecated components, security vulnerabilities or breaking changes [47]. These dependency issues also arise with reusable Actions, underscoring the critical need for Action developers and users to adopt robust dependency management and version control practices, policies, and tools, similar to those employed for reusable OSS libraries [47]. Several such tools already exist, such as GitHub's *Dependabot* or the third-party *Renovate* tool for automating dependency updates, supporting automatic workflow updates to keep the used Actions up-to-date with their latest versions.[22] Tools like OSSF Scorecard[23] and StepSecurity[24] further help to harden workflow security, for example through the mechanism of *Action pinning*. It replaces the insecure use of git tags for referring to Action versions [52] by the more secure practice of directly referencing a git commit hash. Note that this still does not guarantee full protection, as highlighted by security analyses.[25] It is important to raise awareness and increase the take-up of such tools in order to further reduce workflow outdatedness and increase their security posture [20–22]. This need for proactive dependency management and version control practices, policies, and tools is further highlighted by one respondent: *"The biggest issue I used to face is changes introduced by people relabelling GitHub Action tags. For example, it is common to have some/action@v2, and the v2 tag is moved every time a new v2 update is available. This, in theory, should not introduce breaking changes if they follow semantic versioning, but in practice I have occasionally had unexpected changes in behaviour. I now always pin actions to specific hashes some/action@hash # v2.1.5, and use Renovate to manage updates, allowing me to inspect changelogs, and allowing me to easily revert a change should it not work."*

Our findings on the challenges of adopting reusable Actions are compounded by the high prevalence of certain workflow failure types. Analysis of failed workflows across the GitHub ecosystem reveals that test, compilation, and configuration issues are the most frequent causes of failure [71]. Given that reusable Actions introduce external code and complex configurations into a workflow, these findings underscore how issues related to dependency management and component integration are often manifested as configuration and compilation failures, contributing significantly to developer debugging effort.

---

RECOMMENDATIONS:

- Providers of reusable Actions should (1) prioritise the non-functional characteristics of Actions, such as reliability, security, and maintainability; (2) adopt or even impose robust change control practices, like semantic versioning, to ensure that breaking changes and deprecated Actions are clearly communicated and managed effectively across GitHub and its Marketplace; (3) provide clear guidelines and best practices for Action developers on how to implement and maintain these non-functional aspects, including security hardening techniques and comprehensive testing strategies.

- Researchers should define metrics and measurement frameworks for assessing the non-functional qualities of Actions and evaluating aspects like security posture, code quality, documentation completeness, and adherence to best practices.

- Workflow practitioners should be aware of the security risks associated with relying on reusable components, especially Actions. It is crucial to employ robust security tools to thoroughly assess these potential vulnerabilities. This careful attention is particularly vital given that Actions are increasingly becoming a target for supply chain attacks.

---

[22]https://docs.github.com/en/code-security/dependabot/working-with-dependabot/keeping-your-actions-up-to-date-with-dependabot
[23]https://github.com/ossf/scorecard
[24]https://app.stepsecurity.io
[25]https://www.paloaltonetworks.com/blog/cloud-security/unpinnable-actions-github-security

## 6 Threats to validity

We report on the four types of validity threats pertaining to our research, following the recommendations of Wohlin et al. [67].

**Construct validity** examines the extent to which the survey questions accurately measure the intended theoretical constructs. We addressed potential threats to construct validity by following established survey design guidelines [41] throughout the questionnaire development process. This included several rounds of reviews and collaborative discussions among the co-authors to refine question wording and ensure clarity and relevance to our research goals ($G_1$ and $G_2$). In addition, a pilot survey with experienced GitHub Actions practitioners provided valuable feedback on the questionnaire's clarity and any potential ambiguities, leading to appropriate revisions. A remaining potential threat to construct validity lies in the choice of English for conducting the survey. While English is the predominant language within open-source communities and on GitHub, this choice may have inadvertently influenced the responses or excluded respondents whose mother tongue is not English.

**Internal validity** concerns factors within a study that could influence the observed outcomes, independent of the intended manipulations or measurements. One such threat arises from the selection of repositories and respondents. Our sampling frame focused on GitHub contributors to OSS repositories who demonstrated recent and significant activity in maintaining GitHub Actions workflow files. As such, our sample may not fully represent all workflow maintainers, particularly those who are less active or less experienced in using GitHub Actions. This selection bias could influence the findings, as more experienced or engaged users may have different perspectives and practices compared to less active users.

**Conclusion validity** addresses the degree to which the conclusions drawn from our analysis are reasonable and supported by the data. One potential threat stems from the conversion of ordinal data from the four-point Likert scales into numerical values for statistical analysis. We followed the common practice of assigning linear numerical values to the ordered labels (e.g., *Unimportant* = 0, *Slightly Important* = 1, *Moderately Important* = 2, *Very Important* = 3), a widely accepted method in the analysis of Likert-scale data. Alternative assignment strategies could yield different results, raising a conclusion validity concerning the interpretation of statistical measures that assume interval-level data [50]. Another decision that could impact conclusion validity relates to the grouping of some Likert-scale responses for certain analyses. For example, our decision to group *never* or *unimportant* on one side and all other options on the other aimed to reduce subjectivity by broadly categorising negative and positive responses. Different groupings could lead to different interpretations. Apart from this, our conclusions are primarily based on descriptive statistics, non-parametric statistical tests (Kruskal-Wallis and Dunn's tests), and a partial ordering derived from these tests. By applying appropriate statistical methods, including the Benjamini-Hochberg correction to control the family-wise error rate due to multiple testing, we aimed to ensure the robustness and reliability of our findings, thus minimising threats to conclusion validity.

**External validity** concerns the extent to which the findings of this study can be generalised or extended beyond the specific research boundaries. OSS contributors are known to come predominantly from Europe, the US, and Canada. Since we did not collect any geographical data from respondents (due to privacy regulations), our findings should therefore not be generalised outside this specific population. Finally, we cannot generalise the findings to other collaborative coding platforms and/or CI/CD services (such as GitLab and its CI/CD pipelines), due to their specificities.

## 7 Conclusion

This study surveyed 419 practitioners to understand their GitHub Actions workflow development and maintenance practices, aiming to identify common challenges and opportunities for improved automation support. We gained valuable insights into the prevailing automation and reuse practices amongst GitHub Actions workflow practitioners, revealing that the full potential for automation across the CI/CD spectrum is yet to be realised. This is largely due to a concentration on core CI/CD tasks, with aspects like security analysis and health monitoring receiving less attention. To fully leverage GitHub Actions, there is a clear need for increased awareness and improved tooling to reduce the technical complexity associated with manually testing, debugging, and maintaining workflows and their reusable components.

Our investigation into reuse practices highlighted a strong inclination towards employing reusable Actions, indicating the success of GitHub's component-based approach. However, the less frequent adoption of reusable workflows, coupled with the common practice of copy-pasting, particularly from one's own codebase, suggests a preference for familiarity and control, potentially influenced by technical limitations and a lack of awareness and understanding of other reuse mechanisms. The dependency challenges encountered when using Actions also hinder their reuse, potentially pushing practitioners towards more cautious, self-reliant strategies.

The non-functional characteristics deemed most important by workflow maintainers – namely reliability, security, and understandability – underscore the pragmatic concerns of ensuring stable, secure, and easily maintainable automation. The reported difficulties with Action usage directly affect these qualities, emphasising the critical need for improvements in version management, tooling, and community best practices, including addressing the existing limitations in platform support for essential maintenance activities such as testing and debugging.

Overall, whilst GitHub Actions provides a powerful platform for workflow automation and reuse, fully realising its potential requires addressing the observed imbalances in automation adoption and the complexities surrounding reuse practices. Future efforts should concentrate on making advanced automation features more accessible and user-friendly, leveraging AI-based solutions (such as GitHub Copilot and its coding agent,[26] and other similar tools) to significantly ease workflow creation and maintenance, particularly for developers for whom this is not their core business. These efforts should also focus on improving the discoverability and dependability of reusable components, and fostering a deeper understanding of the advantages and best practices associated with reuse mechanisms in workflows. This will empower practitioners to leverage automation and reuse more effectively, driving increased efficiency, improved software quality, and a reduced maintenance burden in a more sustainable ecosystem.

### Acknowledgments

This research is supported by the Fonds de la Recherche Scientifique - FNRS under grant numbers T.0149.22, F.4515.23 and J.0147.24.

---

[26]https://docs.github.com/en/copilot